\documentclass[usenatbib,useAMS]{mn2e}
\usepackage{natbib,times}
\usepackage{longtable}
\usepackage{amssymb}
\usepackage{rotating}
\usepackage{epsfig}
\usepackage{amsmath}
\usepackage{graphicx}
\usepackage{makecell}
\def\hMpc{~$h^{-1}{\rm Mpc}$}
\def\kms{km~s$^{-1}$}
\def\HI{H\,{\sc i}}

\begin{document}

\title[2MTF IV. Bulk flow]
{2MTF IV. A bulk flow measurement of the local Universe}

\author[T. Hong et al.]
{Tao~Hong~$^{1,2,3}$\thanks{E-mail: bartonhongtao@gmail.com}, 
Christopher~M.~Springob~$^{2,3,4}$, 
Lister~Staveley-Smith~$^{2,3}$,
\newauthor
Morag~I.~Scrimgeour~$^{5,6,2,3}$, 
Karen~L.~Masters~$^{7,8}$, 
Lucas~M.~Macri~$^{9}$, 
\newauthor
B\"arbel~S.~Koribalski~$^{10}$, 
D.~Heath~Jones~$^{11}$, 
and Tom~H.~Jarrett~$^{12}$
\\
%\vspace*{1cm}
%\scriptsize{
$^{1}$National Astronomical Observatories, Chinese Academy
  of Sciences, 20A Datun Road, Chaoyang District, Beijing 100012,
  China.\\
$^{2}$International Centre for Radio Astronomy Research,
  M468, University of Western Australia, Crawley, 35 Stirling Highway, WA 6009, Australia\\
$^{3}$ARC Centre of Excellence for All-sky Astrophysics
  (CAASTRO)\\
$^{4}$  Australian Astronomical Observatory, PO Box 915, North Ryde, NSW
1670, Australia\\ 
$^{5}$ Department of Physics and Astronomy, University of Waterloo, Waterloo, ON, N2L 3G1, Canada\\
$^{6}$ Perimeter Institute for Theoretical Physics, 31 Caroline St. N., Waterloo, ON, N2L 2Y5, Canada\\
$^{7}$Institute for Cosmology and Gravitation, University 
of Portsmouth, Dennis Sciama Building, Burnaby Road, Portsmouth 
PO1 3FX\\
$^{8}$South East Physics Network (www.sepnet.ac.uk)\\
%$^{9}$Harvard-Smithsonian Center for Astrophysics, 60 Garden Street, Cambridge, MA 02138, USA\\
$^{9}$George P. and Cynthia Woods Mitchell Institute for Fundamental Physics and 
Astronomy, Department of Physics and Astronomy, Texas A\&M University, \\4242 TAMU, 
College Station, TX 77843, USA\\ 
$^{10}$CSIRO Astronomy \& Space Science, Australia Telescope National 
Facility, PO Box 76, Epping, NSW 1710, Australia\\ 
$^{11}$School of Physics, Monash University, Clayton, VIC 3800, Australia\\ 
$^{12}$Astronomy Department, University of Cape Town, Private Bag X3. 
Rondebosch 7701, Republic of South Africa\\
}

\date{Accepted  ... Received  ...}

\pagerange{\pageref{firstpage}--\pageref{lastpage}} \pubyear{}

\maketitle
\label{firstpage}
\begin{abstract}
Using the 2MASS near-infrared photometry and high signal-to-noise \HI\ 
21-cm data from the Arecibo, Green Bank, Nancay, and Parkes 
telescopes, we calculate the redshift-independent distances and
peculiar velocities of 2,018 bright inclined spiral galaxies over the
whole sky.  This project is part of the 2MASS Tully-Fisher survey
(2MTF), aiming to map the galaxy peculiar velocity field within
100\hMpc, with an all-sky coverage apart from Galactic latitudes
$|b|<5^{\circ}$.  A $\chi^2$ minimization method was adopted to
analyze the Tully-Fisher peculiar velocity field in J, H and K bands,
using a Gaussian filter.  We combine information from the three
wavebands, to provide bulk flow measurements of $310.9 \pm 33.9$
\kms, $280.8 \pm 25.0$ \kms, and $292.3 \pm 27.8$ \kms\ at
depths of 20\hMpc, 30\hMpc\ and 40\hMpc, respectively. Each of these
bulk flow vectors points in a direction similar to those found by
previous measurements. At each of the three depths, the bulk flow
magnitude is consistent with predictions made by the $\Lambda$CDM
model at the $1\sigma$ level. The maximum likelihood and minimum
variance method were also used to analyze the 2MTF samples, giving similar results.

\end{abstract}
\begin{keywords}
galaxies: distances and redshifts --- galaxies: spiral 
--- radio emission lines  --- catalogs --- surveys
\end{keywords}

\section{Introduction}
Galaxy redshifts exhibit deviations from Hubble's law known as `peculiar
velocities', which are induced by the gravitational attraction of all
surrounding matter.  Peculiar velocities thus provide a means to trace
the overall matter density field and detect all gravitating matter
(both visible and dark).

\begin{figure*}
\centering
\includegraphics[width=0.9\columnwidth, angle=-90]{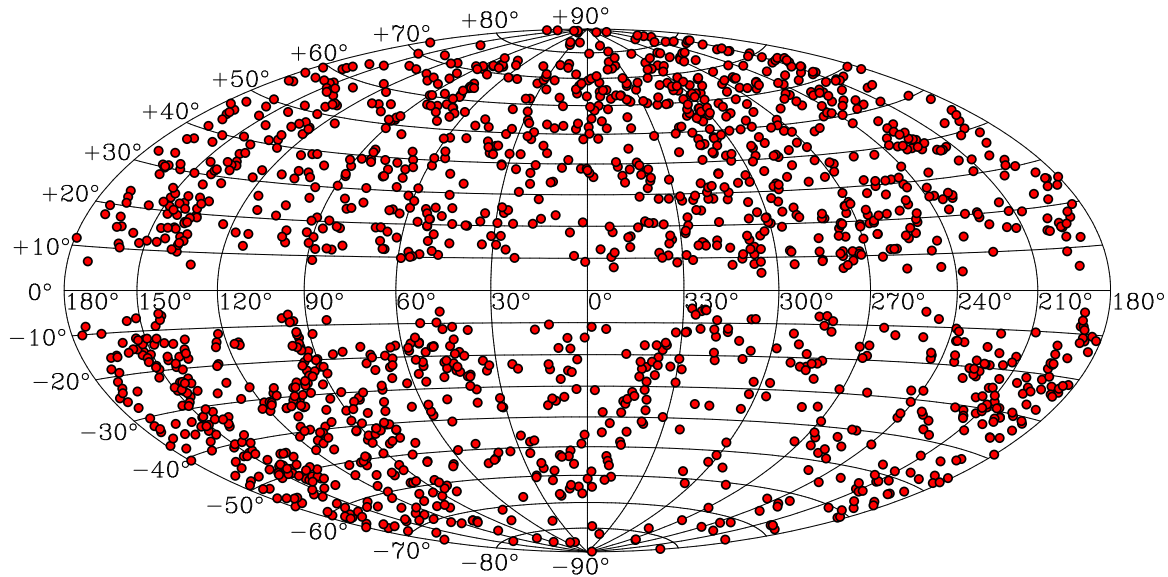}
\caption{Sky coverage of the 2,018 2MTF galaxies, in Galactic
  coordinates with an Aitoff projection. }
\label{fig:sky}
\end{figure*}
Peculiar velocities can be measured using redshift-independent
distance indicators.  Several such indicators have been used, including Type Ia SNe \citep{Phillips1993}, the
Tully-Fisher relation \citep[TF,][]{Tully1977} and the Fundamental
Plane relation \citep[FP,][]{Djorgovski1987,Dressler1987}.  The
largest peculiar velocity surveys make use of either the FP or TF
relations.  One limitation of these, and many other, redshift
independent distance indicators, however, is that they rely on optical
photometry or spectroscopy.  This means that the optical extinction effects of our
Galaxy limit the sky coverage of such surveys in the so-called Zone of Avoidance (ZoA).

The 2MASS Tully-Fisher Survey \citep[2MTF,][]{Masters2008, Hong2013, Masters2014}
is an all-sky Tully-Fisher survey aiming to measure the
redshift-independent distances of nearby bright spiral galaxies. All
galaxies in the 2MTF sample were selected from the 2MASS Redshift
Survey \citep[2MRS,][]{Huchra2012}, which is an extended redshift
survey based on the 2 Micron All-Sky Survey Extended Source Catalog
\citep[2MASS XSC,][]{Jarrett2000}.  By using the near-infrared
photometric data and high quality 21-cm \HI\ data from the Green Bank
Telescope (GBT), Parkes radio telescope, Arecibo telescope, and other
archival \HI\ catalogs, 2MTF provides sky coverage down to Galactic
latitude $|b| = 5 ^\circ$, which significantly reduces the area of the
ZoA relative to previous Tully-Fisher surveys.

One parameter of the peculiar velocity field that benefits from the
improved sky coverage of 2MTF is the dipole, or `bulk flow'.  In past
measurements of the bulk flow of the local universe, authors have
largely agreed on the direction of the flow, but disagree on its amplitude
\citep{Hudson2004,Feldman2010,Dai2011,Ma2012,
  Ma2013,Rathaus2013,Ma2014}. \citet{Watkins2009} analyzed a peculiar
velocity sample of 4,481 galaxies with a ``Minimum Variance" method,
within a Gaussian window of radius 50\hMpc.  They found a bulk flow
with amplitude $407 \pm 81$~\kms, which they claim is inconsistent with the
$\Lambda$CDM model at $>$98\% confidence level.

On the other hand, other studies have shown a bulk flow amplitude
agreeing with the $\Lambda$CDM model. By estimating the bulk flow of
the SFI++ Tully-Fisher sample \citep{Masters2006,Springob2007},
\citet{Nusser2011} derived a bulk flow of $333 \pm 38$~\kms\ at a
depth of 40\hMpc, which is consistent with the $\Lambda$CDM model.
\citet{Turnbull2012} adopted the bulk flow estimation on a Type Ia SNe
sample with 245 peculiar velocity measurements, finding a bulk flow of
$249 \pm 76$~\kms, which is also consistent with the expectation
from $\Lambda$CDM.

In this paper, we use 2MTF data to measure the bulk flow in the
local Universe. The relatively even sky coverage and uniform source
selection criteria make 2MTF a good sample for bulk flow analysis. The
data collection is described in
Section~\ref{sec:data}. Section~\ref{sec:tf_pv} introduces the
calculations of Tully-Fisher distances and peculiar velocities. 
A $\chi^2$ minimization method was adopted to analyze the peculiar
velocity field, which is presented in Section~\ref{sec:bulk}.  We
compared our measurements with the $\Lambda$CDM predictions and
previous measurements in Section~\ref{sec:compare}.  A summary is
provided in Section~\ref{sec:conclu}.

Throughout this paper, we adopt a spatially flat cosmology with $\Omega_\mathrm{m} =
0.27$, $\Omega_\Lambda = 0.73$, $n_\mathrm{s} = 0.96$ \citep[WMAP-7yr,][]{Larson2011}. 
% All distances in
%this paper are calculated and published in units of \hMpc, \textbf{where $h$ 
%is the Hubble parameter in units of km s$^{-1}$ Mpc$^{-1}$}.
All distances in this paper are calculated and published in units of \hMpc, 
where the Hubble constant is given by $H_0=100 h$ km s$^{-1}$ Mpc$^{-1}$. The 
observational results in this paper are independent of the actual value of $h$, though 
we compare with a $\Lambda$CDM model in Section~\ref{sec:compare} which assumes the WMAP-7yr value 
of $h=0.71$.

\section{Datasets}
\label{sec:data}
To build a high quality peculiar velocity sample, all 2MTF target
galaxies are selected from 2MRS using the following criteria: 
total K magnitudes K $ < 11.25$~mag, $cz<10,000$~\kms, and
axis ratio $b/a < 0.5$.  There are $\sim 6,000$ galaxies that meet
these criteria, but many of them are faint in \HI, so observing the 
entire sample would be very expensive in terms of telescope 
time. We have combined 
archival \HI\ data with observations from the Arecibo Legacy Fast ALFA survey
\citep[ALFALFA,][]{Giovanelli2005} and new observations made with the
GBT and Parkes radio telescope.  The new observations preferentially
targeted late-type spirals deemed likely to be \HI\ rich, but the
archival data includes all spiral subclasses.  Accounting for galaxies
eliminated from the sample because of confusion, marginal SNR, 
non-detections, and other problematic cases, as well as cluster
galaxies reserved for use in the 2MTF template \citep{Masters2008},
current peculiar velocity sample from 2MTF catalog contains 2,018 galaxies. As shown in
Figure~\ref{fig:sky}, the catalog provides uniform sky coverage down
to Galactic latitude $|b|=5^\circ$, which makes it a good sample for
bulk flow and large-scale structure analysis.

We note that the sample of 2,018 galaxies discussed here is {\it
  separate} from the 888 cluster galaxies used to fit the template
relation for 2MTF in \citet{Masters2008}.  The relationship between
that template sample and the field sample is discussed further in
Section~\ref{sec:template}.

\subsection{Photometric Data}
\label{sec:photo}
All 2MTF photometric data are obtained from the 2MRS catalog. 2MRS is
an all-sky redshift survey based on 2MASS.  It provides more than
43,000 redshifts of 2MASS galaxies, with K $ < 11.75$~mag and $|b| \geq
5^\circ$. The photometric quantities which are adopted for the
Tully-Fisher calculations are the total magnitudes in J, H and K
bands, the 2MASS co-added axis-ratio $b/a$ \citep[the axis-ratio of
  J+H+K image at the 3$\sigma$ isophote; for more details
  see][]{Jarrett2000} and the morphological type code $T$.

\begin{figure}
\centering
\includegraphics[width=0.95\columnwidth, angle=-90]{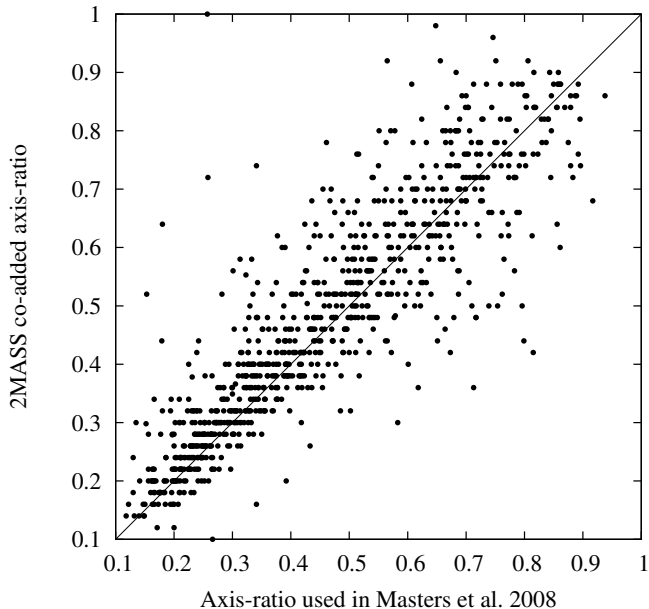}
\caption{The comparison between the axis-ratios adopted by
  \citet{Masters2008} and the 2MASS co-added axis-ratios.  The solid
  line indicates equality. 888 template galaxies used by
  \citet{Masters2008} are plotted here. The scatter of these two
  axis-ratios is 0.096.}
\label{fig:inc}
\end{figure}

\citet{Masters2008} built the 2MTF Tully-Fisher template using the
axis-ratio in I-band and J-band. However, the I-band axis-ratios are
only available for a small fraction of the 2MTF non-template galaxies.
Thus, to make the 2MTF non-template sample more uniform, we chose the
2MASS co-added axis-ratio as our preferred data. As shown in
Figure~\ref{fig:inc}, we took the 888 2MTF template galaxies as a
comparison sample and compared the 2MASS co-added axis-ratios with the
axis-ratios used in the 2MTF template. A dispersion was found between
these two parameters ($\sim 0.096$), which we expect will introduce a
small scatter into the final Tully-Fisher relation, but no significant
bias was detected in this comparison.

Galaxy internal dust extinction and $k$-correction were done following
\citet{Masters2008}. The total magnitudes provided by the 2MRS catalog
have already been corrected for Galactic extinction, so no additional
such correction was applied here.

\subsection{\HI\ Rotation Widths}

\subsubsection{Archival Data}

Before making our new \HI\ observations, we collected high-quality \HI\ 
width measurements from the literature. The primary source of 2MTF
archival data is the Cornell \HI\ digital archive
\citep{Springob2005}. The Cornell \HI\ digital archive provides about
9,000 \HI\ measurements in the local Universe ($-200 < cz < 28,000$
\kms) observed by single-dish radio telescopes.  When cross-matching
this \HI\ dataset with the 2MRS catalog, we include only the galaxies
with quality codes suggesting suitability of the measured width for TF
studies, G (Good) and F (Fair) in the nomenclature of
\citet{Springob2005}.  (See Section 4 of that paper for details.)
1,038 well-measured galaxies in this dataset which meet the 2MTF
selection criteria were taken into the final 2MTF catalog. The \HI\ 
widths provided by \citet{Springob2005} were corrected for redshift
stretch, instrumental effects and smoothing, we have added the
turbulence correction and viewing angle correction to make the
corrected widths suitable for the Tully-Fisher calculations.

Besides the Cornell \HI\ digital archive, we also collected \HI\ data
from \citet{Theureau1998, Theureau2005, Theureau2007, Mathewson1992}
and Nancay observed galaxies in Table A.1 of \citet{Paturel2003}.  The
raw observed widths taken from these sources were corrected for the
aforementioned observational effects following \citet{Hong2013}.

\subsubsection{GBT and Parkes Observations}
In addition to the archival \HI\ data, new observations were conducted
by the Green Bank Telescope (GBT) and the Parkes radio telescope
between 2006 to 2012.

1,193 2MTF target galaxies in the region of $\delta > -40^\circ$ were
observed by the GBT \citep{Masters2014} in position-switched
mode, with the spectrometer set at 9 level sampling with 8192
channels.  After smoothing, the final velocity resolution was 5.15
\kms.  The \HI\ line was detected in 727 galaxies, with 483 of them
considered good enough to be included into the 2MTF catalog.

The Parkes radio telescope was used to observe the targets in the
declination range $\delta \leq -40^\circ$ \citep{Hong2013}. We
observed 305 galaxies which did not already have high quality \HI\ 
measurements in the literature and obtained 152 well detected \HI\ 
widths (signal-to-noise ratio SNR $>$ 5). Beam-switching mode with 7
high-efficiency central beams was used in the Parkes observations.
The multibeam correlator produced raw spectra with a velocity
resolution $\Delta v \sim 1.6$ \kms , while we measured \HI\ widths and
other parameters on the Hanning-smoothed spectra with a velocity
resolution of $\sim 3.3$ \kms.

All newly-observed spectra from the GBT and Parkes telescope were
measured by the same IDL routine \textit{awv\_fit.pro}, which based on 
the algorithm used by \citet{Springob2005}. \HI\ line widths were measured by five
different methods \citep[see][Section 2.1.2]{Hong2013}.  We chose
$W_{F50}$ as our preferred width algorithm. This fitting method can
provide accurate width measurements for low S/N spectra.  It was
adopted as the preferred method by the Cornell \HI\ digital archive and
the ALFALFA blind \HI\ survey.

\subsubsection{ALFALFA Data}

The GBT observations did not cover the entirety of the sky north of
$\delta = -40^\circ$.  7,000 deg$^2$ of the northern (high Galactic
latitude) sky were observed by the ALFALFA survey. 
In this region of the sky, we
rely on the ALFALFA observations, rather than making our own GBT
observations.  In this region, ALFALFA is expected to detect more than 30,000
\HI\ sources with \HI\ mass between $10^6$ M$_\odot$ and $10^{10.8}$
M$_\odot$ out to z $\sim$ 0.06.  The ALFALFA dataset has not been
fully released yet.  \citet{Haynes2011} published the $\alpha. 40$
catalog which covers about $40\%$ of the final ALFALFA sky.  However,
the ALFALFA team has provided us with an updated catalog of \HI\ widths,
current as of October 2013.  The new catalog covers about 66\% of the
ALFALFA sky. From this database, the 2MTF catalog obtained 576 \HI\ 
widths.  When the final ALFALFA catalog is released, we will
correspondingly update the 2MTF catalog, making use of the complete
dataset.

\subsection{Data collecting limit of Tully-Fisher sample}
\begin{figure}
\centering
\includegraphics[width=0.65\columnwidth, angle=-90]{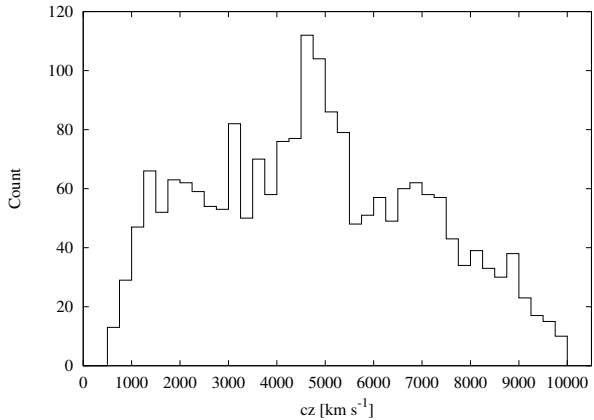}
\caption{The redshift
  distribution of the 2MTF sample in the CMB frame.}
\label{fig:redshift}
\end{figure}

The initial 2MTF catalog includes 2,708 galaxies.  An additional
cutoff was employed to improve the data quality. Only the well-measured 
galaxies were included in the 2MTF peculiar velocity sample.  That is, we
include those galaxies with $cz \geq 600$ \kms, relative \HI\ width
error $\epsilon_\mathrm{w}/w_{\mathrm{HI}} \leq 10\%$ and \HI\ spectrum signal-to-noise
ratio SNR $\geq 5$. 141 galaxies in the template sample of
\citet{Masters2008} were excluded. The current 2MTF peculiar velocity sample has 2,018
galaxies with good enough data quality for the Tully-Fisher distance
calculations and further cosmological analysis. The sky coverage and
redshift distribution of the 2MTF peculiar velocity sample are plotted in
Figure~\ref{fig:sky} and Figure~\ref{fig:redshift} respectively.

\section{Derivation of Tully-Fisher distances and Peculiar velocities}
\label{sec:tf_pv}
\subsection{Template Relation}
\label{sec:template}
Using 888 spiral galaxies in 31 nearby clusters, \citet{Masters2008}
built Tully-Fisher template relations in the 2MASS J, H and K bands,
following the approach taken by \citet{Masters2006} for SFI++, which
in turn follows the approach of \citet{Giovanelli1997} for SCI.  The
authors found a dependence of the TF relation on galaxy morphology,
with later spirals presenting a steeper slope and a dimmer zero point
on the relation than earlier type spirals.  This morphological
dependence occurs in all three 2MASS wavebands. The final template
relations were corrected to the Sc type in the three bands, as was
done for in the I-band relation examined by \citet{Masters2006}.

We use the revised version of the \cite{Masters2008} Tully-Fisher
relation as our template to calculate the distances of the 2MTF
galaxies:
\begin{equation}
\begin{aligned}
M_K - 5\log h=-22.188-10.74(\log W -2.5),\\
M_H - 5\log h=-21.951-10.65(\log W -2.5),\\
M_J - 5\log h=-21.370-10.61(\log W -2.5),
\end{aligned}
\label{eq:tf}
\end{equation}
where W is the corrected \HI\ width in units of \kms, and $M_K, M_H$,
and $M_J$ are the absolute magnitudes in the three bands
separately.

\subsection{TF distance and peculiar velocity calculations}
\label{sec:tf}

Since the components of the peculiar velocity uncertainties are log-normal, 
our bulk flow analysis is done
in logarithmic space. The main parameters used in the fitting
processes are $\log(d_{\mathrm{TF}})$ and its corresponding logarithmic error.
More precisely, rather than express the peculiar velocity in linear
units, we work with the closely related logarithmic quantity
$\log(d_{z}/d_{\mathrm{TF}})$, which is the logarithm of the ratio between the
galaxy's redshift distance in the CMB frame and its Tully-Fisher-derived 
true distance.  At this stage, we express this as
$\log(d_{z}/d_{\mathrm{TF}}^{*})$, which represents the logarithmic distance
ratio before the Malmquist bias correction has been applied (see
Section 3.4.)  We can then express this as:
\begin{equation}
\label{eq:distance}
\log\left({\frac{d_{z}}{d_{\mathrm{TF}}^{*}}}\right) = {\frac{-\Delta M}{5}},
\end{equation}
where $\Delta M = M_{\mathrm{obs}}-M(W)$ is the difference between the
corrected absolute magnitude $M_{\mathrm{obs}}$ (calculated using the redshift distance of the galaxy) 
and the magnitude generated from the TF
template relation $M(W)$ \citep[for more details on corrected absolute
  magnitudes, see Equation~7 from][]{Masters2008}.

The errors on the logarithmic distance ratio are the sum in quadrature
of the \HI\ width error, near-infrared magnitude error, inclination
error, and the intrinsic error of the Tully-Fisher relation.  Instead
of using the intrinsic error estimates reported by
\citet{Masters2008}, we adopted new intrinsic error terms:
\begin{equation}
\begin{aligned}
\epsilon_{\mathrm{int}, K}=0.44-0.66(\log W -2.5),\\
\epsilon_{\mathrm{int}, H}=0.45-0.95(\log W -2.5),\\
\epsilon_{\mathrm{int}, J}=0.46-0.77(\log W -2.5).
\end{aligned}
\label{eq:intrinsic}
\end{equation}
These relations were derived in magnitude units.  To transform the
errors into the logarithmic units used for $\log(d_{z}/d_{\mathrm{TF}}^{*})$,
one must divide the $\epsilon_{\mathrm{int}}$ values by 5.  We describe the
details of the new intrinsic error term estimation in
Appendix~\ref{sec:intrinsic}.

As shown in Figure~\ref{fig:inc}, the difference between the two
sources of inclination introduces a mean scatter of 0.096, so we
adopted a uniform scatter $\sigma_{\mathrm{inc}} = 0.1$ as the inclination
error. 
The inclination uncertainty then introduces errors into the Tully-Fisher distances. For a galaxy with inclination $b/a$, 
we calculated the corrected \HI\ width with $b/a \pm 0.1$, and got the 
corrected widths $W_{\mathrm{high}}$ and $W_{\mathrm{low}}$ respectively. 
Half of the difference between the two widths, 
$\Delta_{\mathrm{wid, inc}} = 0.5(W_{\mathrm{high}} - W_{\mathrm{low}})$, was then adopted 
as the error introduced by the inclination into the \HI\ corrected width. 
Finally, the error introduced to the Tully-Fisher distances by inclination uncertainty, 
$\epsilon_{\mathrm{inc}}$, was calculated using the same approach undertaken to calculate the contribution from the HI width uncertainty, $\epsilon_{\mathrm{wid}}$.

\begin{figure*}
\centering
\includegraphics[width=0.6\columnwidth, angle=-90]{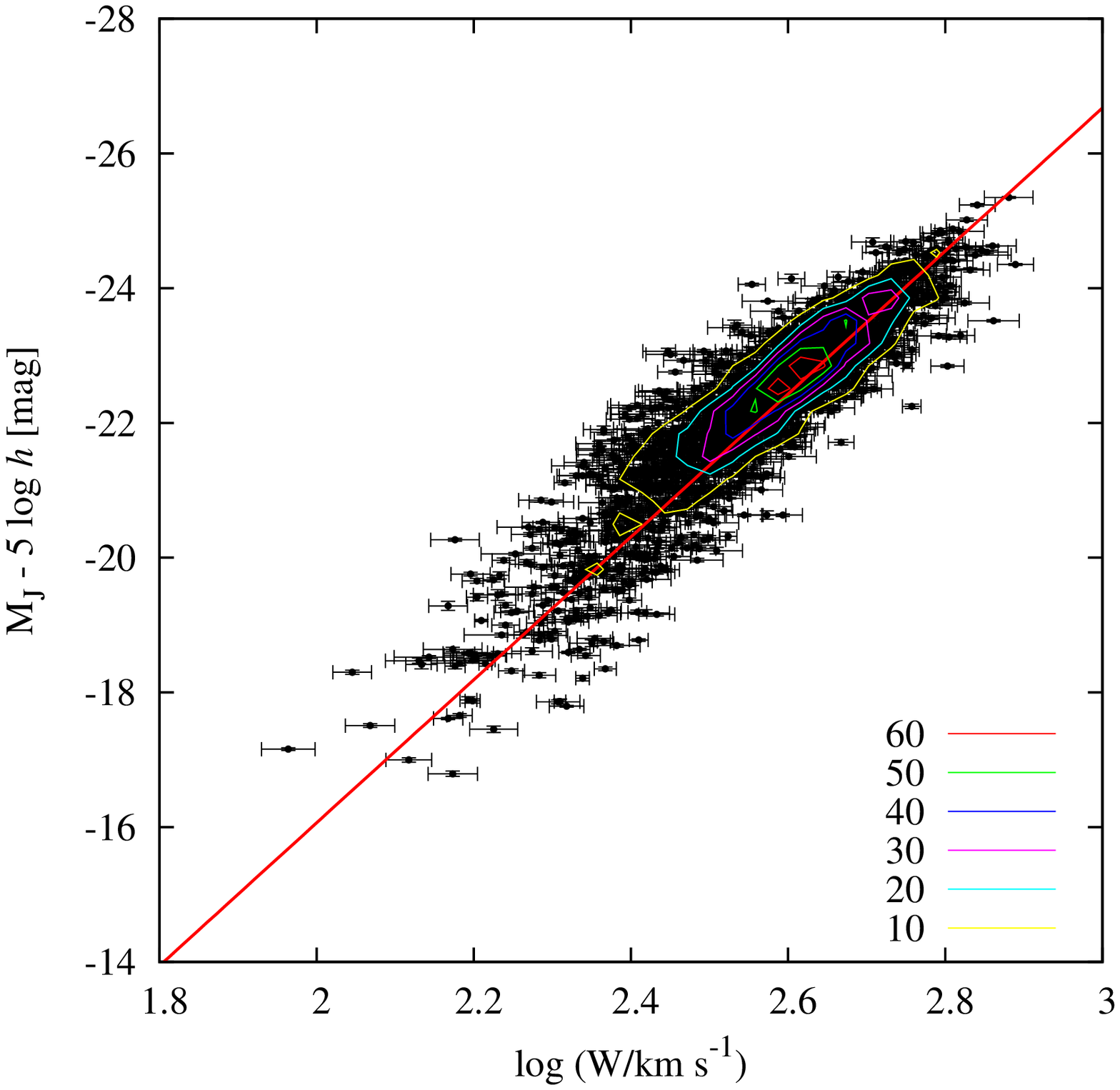}\hspace*{\fill}
\includegraphics[width=0.6\columnwidth, angle=-90]{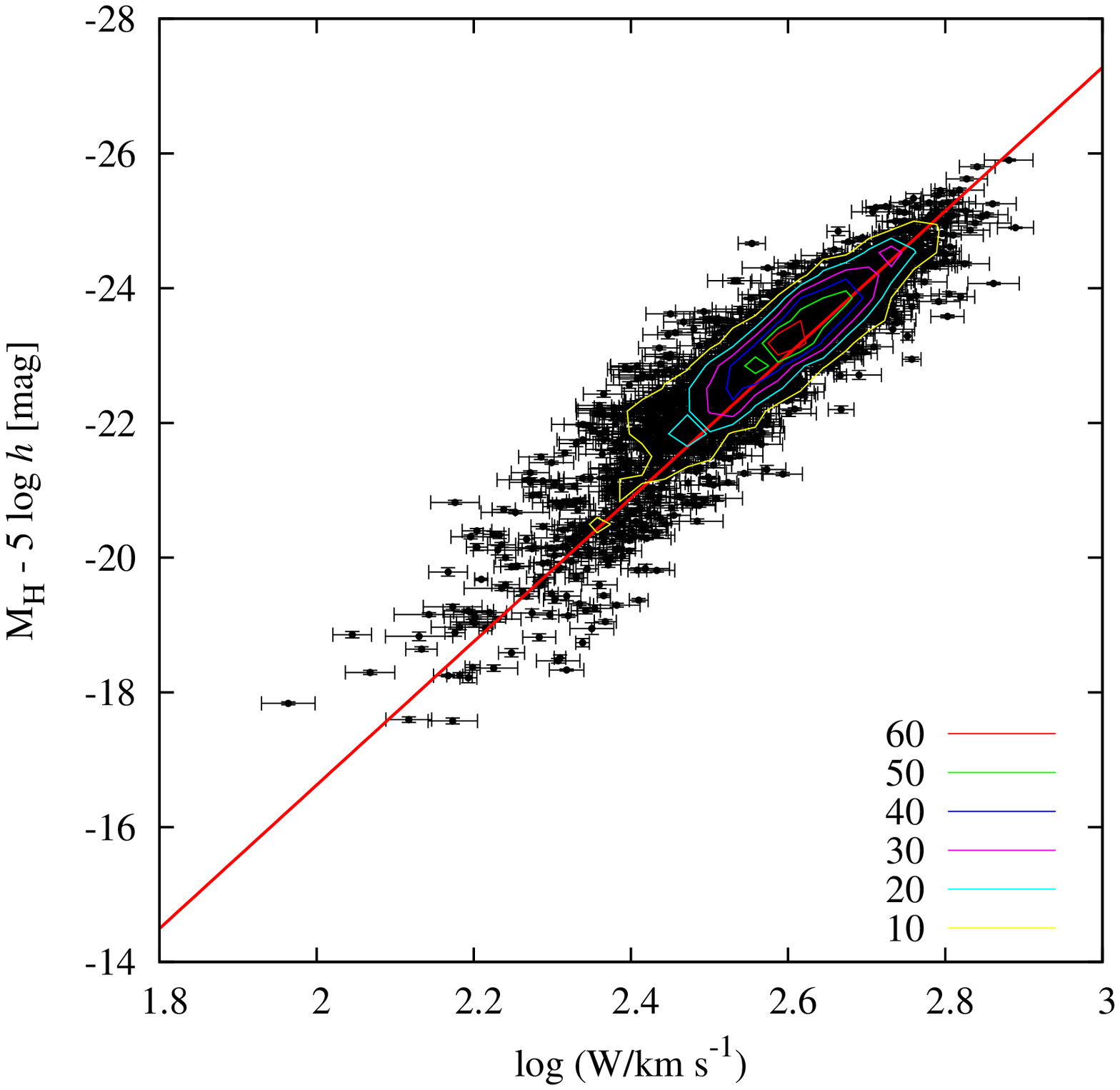}\hspace*{\fill}
\includegraphics[width=0.6\columnwidth, angle=-90]{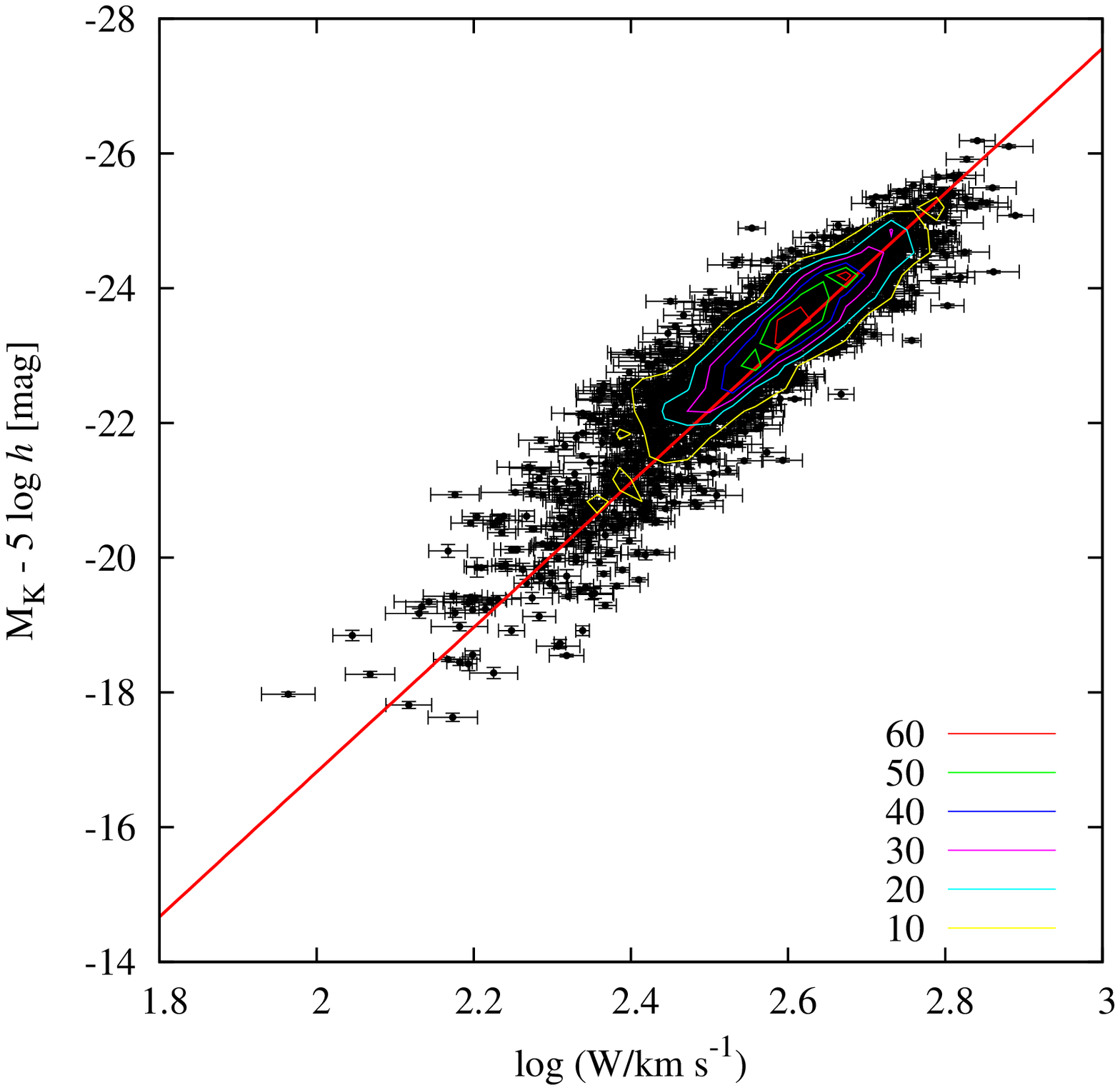}\hspace*{\fill}
\caption{Tully-Fisher relations for 2MTF galaxies in the J, H and K bands (left to right). 
 The red solid lines are the TF template
  relations in the three bands. By making a $50\times50$ grid on the
  Tully-Fisher relation surface $\log W - M $, we counted the number
  of galaxies falling in every grid point, and took these counts as
  the number density of the Tully-Fisher plot, which are indicated by
  the color contours.}
\label{fig:tf_plot}
\end{figure*}

The Tully-Fisher relations in the three bands are shown in
Figure~\ref{fig:tf_plot}, along with the template relations.  We also
show the observational errors for both the \HI\ width and NIR
magnitudes.

\subsection{Member galaxies of galaxy groups}
We cross-matched the 2MTF galaxies with the group list identified by
\citet{Crook2007}. 55 galaxy groups have more than one member in the
2MTF sample. We assume all member galaxies in a given group have the
same redshift. That is, when doing the calculation
in Equation~\ref{eq:distance}, all member galaxies in a group are
assigned the group redshift $cz_{\mathrm{group}}$, while the magnitude offset
$\Delta M$ was still calculated separately for each galaxy.

%For a group with $N$ member galaxies in the 2MTF sample, the distance
%of the group is:
%\begin{equation}
%d_{TF, group}=\frac{\sum_{i=1}^{N}d_{TF, i}/\epsilon_i}{\sum_{i=1}^{N}1/\epsilon_i}, 
%\end{equation}
%where $\epsilon_i$ is the error of the Tully-Fisher distance
%calculated by the method in Section~\ref{sec:tf}. The error of the
%group Tully-Fisher distance is:
%\begin{equation}
%\epsilon_{group}=\frac{\sqrt{N}}{\sum_{i=1}^{N}1/\epsilon_i}.
%\end{equation}

\subsection{Malmquist bias correction}
\label{sec:malm}
The term `Malmquist bias' describes a set of biases originating from
the spatial distribution of objects. There are two types of biases
that one may consider.  Inhomogeneous Malmquist bias arises from local
density variations along the line of sight, and is much more pronounced when one is working in real  space.  This is because, as explained by \citet{Strauss1995}, the large distance errors cause the observer to measure galaxy distances scattered away from overdense regions in real space.  In contrast, the much smaller redshift errors mean that this effect is insignificant in redshift space.  While some other TF catalogs, such as SFI++, included galaxy distances in real space, the fact that we operate in redshift space means that inhomogeneous Malmquist bias is negligible.  However,
we must account for the second type of Malmquist bias, homogeneous
Malmquist bias.

Homogeneous Malmquist bias comes about as a consequence of the
selection effects of the survey, which cause galaxies to be
preferentially included or excluded from the survey, depending on
their distance.  Ideally, the survey selection function is a known
analytical function, allowing for a relatively straightforward
correction for selection effects.  However, galaxy peculiar velocity
surveys often have complex selection functions, requiring ad hoc
approximations in the application of Malmquist bias corrections
\citep[e.g.,][]{Springob2007}.

In the case of 2MTF, we used homogeneous criteria in determining which
galaxies to observe.  As explained in Section 2, all 2MRS galaxies
with K$ < 11.25$ mag, $cz < 10,000$ km/s, and $b/a<0.5$ that also met
our morphological selection criteria were targeted for inclusion in
the sample.  However, many of the targeted galaxies were not included
in the final sample, because there was no \HI\ detection, the detection
was marginal, or there was some other problem with the spectrum that
prevented us from making an accurate Tully-Fisher distance estimate.

We thus adopt the following procedure for correcting for Malmquist
bias (explained in more detail by Springob et al. in prep.):

1) Using the stepwise maximum likelihood method
\citep{Efstathiou1988}, we derive the K-band luminosity function for
all galaxies in 2MRS that meet our K-band apparent magnitude, Galactic
latitude, morphological, and axis ratio criteria.  For this purpose,
we include galaxies beyond the 10,000 km/s redshift limit, to simplify
the implementation of the luminosity function derivation.  For this
sample, which we designate the `target sample', we fit a Schechter
function \citep{Press1974}, and find 
$M_k * = -23.1$ and $\alpha = -1.10$.  (The stepwise maximum
likelihood method does not determine the normalization of the
luminosity function, but that is irrelevant for our purposes anyway.)
We note that this luminosity function has a steeper faint end slope
than the 2MASS K-band luminosity function derived by
\citet{Kochanek2001}, who find $\alpha = -0.87$.

2) We next assume that the `completeness', which in this case we take
to mean the fraction of the target sample that is included in our
2MTF peculiar velocity catalog for a given apparent magnitude bin, is a simple function
of apparent magnitude, which is the same across the sky in a given
declination range.  We compute this function, simply taking the ratio
of observed galaxies to galaxies in the target sample for K-band
apparent magnitude bins of width 0.25 mag, separately for two sections
of the sky: north and south of $\delta = -40^\circ$.  This divide in
the completeness north and south of $\delta = -40^\circ$ is due to the
fact that the GBT's sky coverage only goes as far south as
$-40^\circ$, and the galaxies south of that declination were only
observed by the somewhat smaller (and therefore less sensitive) Parkes
telescope.

3) Finally, for every galaxy in the 2MTF peculiar velocity sample, we take the
uncorrected $\log(d_z / d_{\mathrm{TF}}^{*})$ value, and the error $\epsilon_d$,
and compute the initial probability distribution of $\log(d_z /
d_{\mathrm{TF}})$ values, assuming that the errors follow a normal distribution
in these logarithmic units.  For each possible value of the
logarithmic distance ratio $\log(d_z / d_{\mathrm{TF},i})$ within $2\sigma$ of
the measured $\log(d_z / d_{\mathrm{TF}}^{*})$, we weight the probability by
$w_i$, where $1/w_i$ is the completeness (as defined in Step 2)
integrated across the entire K-band luminosity function (derived in
Step 1), evaluated at the $\log(d_z / d_{\mathrm{TF},i})$ in question.  Note
that this involves converting the completeness from a function of
apparent magnitude to a function of absolute magnitude, using the
appropriate distance modulus for the distance in question.

From these newly re-weighted probabilities, we calculate the mean
probability-weighted log(distance), as well as the corrected
logarithmic distance ratio error.  This is our Malmquist
bias-corrected logarithmic distance.

\begin{figure}
\centering
\includegraphics[width=0.65\columnwidth, angle=-90]{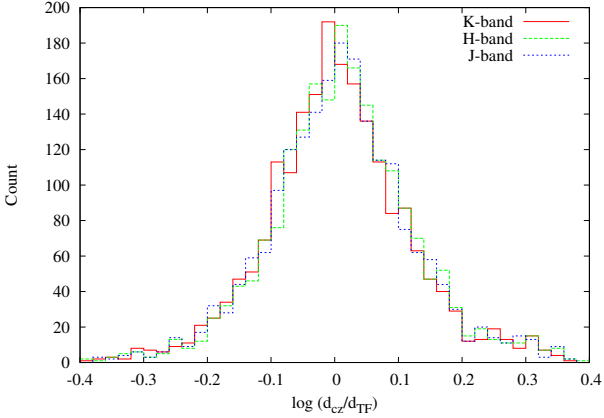}
\caption{The histogram of logarithmic distance ratios $\log
  \dfrac{d_{cz}}{d_{\mathrm{TF}}}$.  The K-band data is shown by the red solid
  line, H-band data by the green dashed line, and the J-band data by
  the blue dotted line.}
\label{fig:ratio}
\end{figure}

\begin{figure}
\centering
\includegraphics[width=0.65\columnwidth, angle=-90]{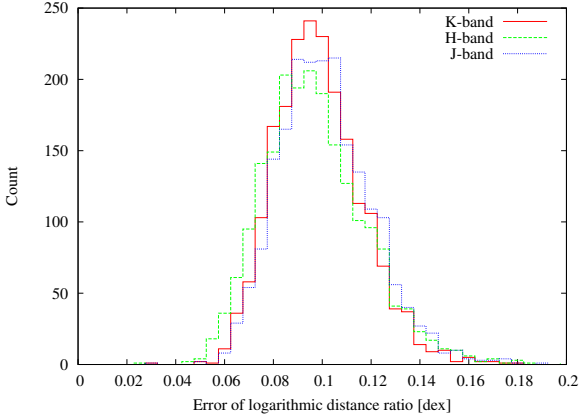}
\caption{The histogram of the error of the logarithmic distance
  ratios, using the same color scheme as in Figure~\ref{fig:ratio}.}
\label{fig:error}
\end{figure}

\begin{figure}
\centering
\includegraphics[width=0.65\columnwidth, angle=-90]{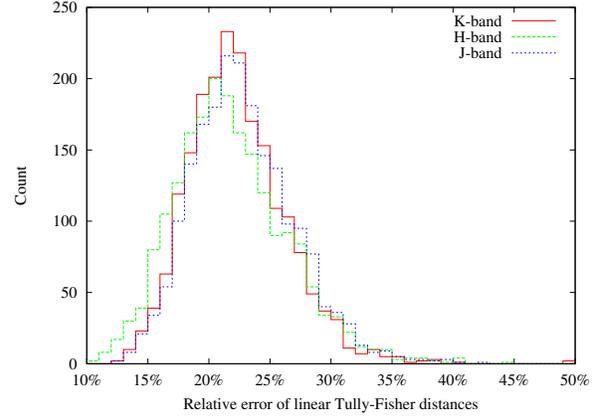}
\caption{The histogram of the relative errors of linear Tully-Fisher distances, 
using the same color scheme as in Figure~\ref{fig:ratio}.}
\label{fig:error_linear}
\end{figure}

The histograms of the logarithmic distance ratios $\log
\dfrac{d_{cz}}{d_{\mathrm{TF}}}$ with the errors are shown in
Figure~\ref{fig:ratio} and Figure~\ref{fig:error} respectively. The
Tully-Fisher distances plotted here are all Malmquist bias-corrected.
A histogram of the relative errors of linear Tully-Fisher distances $d_{\mathrm{TF}}$ is plotted in 
Figure~\ref{fig:error_linear}. The mean errors of the Tully-Fisher distances 
are around $22\%$ in all three bands.

\section{Bulk flow fitting and the results}
\label{sec:bulk}
\subsection{$\chi^2$ minimization method}
\label{sec:method}
We use $\chi^2$ minimization to fit the peculiar velocity field to a
simple bulk flow model. As mentioned in Section~\ref{sec:tf}, the
errors on peculiar velocities are log-normally distributed, so we did
the fitting in logarithmic distance space, similar $\chi^2$ logarithmic methods 
were adopted by \citet{Aaronson1982} and \citet{Staveley-Smith1989}. In the CMB frame, for a
bulk flow velocity $\overrightarrow{V}$, this flow provides a radial
component for each galaxy according to:
\begin{equation}
v_{\mathrm{model}, i} =\overrightarrow{V} \cdot \widehat{r_i},
\end{equation}
where $\widehat{r_i}$ is the unit vector pointing to the galaxy.  We
can also express the model-predicted distance to galaxy $i$ as
\begin{equation}
%d_{model,i}=100 h ^{-1} (cz_i-v_{model,i}) \mathrm{Mpc}.
d_{\mathrm{model},i}=\left(\frac{cz_i-v_{\mathrm{model},i}}{100}\right)h^{-1}\mathrm{Mpc}.
\end{equation}

In our calculation of $\chi^2$, we apply weights which combine the
measurement errors of the logarithmic distance ratios and the weights
generated from both the redshift distribution and the uneven galaxy
number density in different sky areas.  Our $\chi^2$ value is given
by:
\begin{equation}
\label{eq:chi2}
\chi^2 = \sum_{i=1}^{N}\frac{\left[\log(d_{z,i} / d_{\mathrm{model}, i})- \log(d_{z,i} /
    d_{\mathrm{TF}, i})\right]^2 \cdot w^r_i w^d_i} {\sigma_i^2 \cdot \sum_{i=1}^{N}(w^r_i w^d_i)},
\end{equation}
where $\sigma_i$ is the logarithmic distance ratio error of the $i$th
galaxy, $w^r_i$ is the weight arising from the radial
distribution of the sample, and $w^d_i$ is the weight weighting for
the uneven galaxy number density in the northern and southern sky.

The weight $w^r_i$ was designed to make the weighted redshift
distribution of the whole sample match the redshift distribution of an
ideal survey. We adopted the Gaussian density profile following
\citet{Watkins2009}:
\begin{equation}
\label{eq:density}
\rho(r) \varpropto \exp(-r^2/ 2R_I^2),
\end{equation}
with the number distribution
\begin{equation}
\label{eq:number_dist}
n(r) \varpropto r^2 \exp(-r^2/ 2R_I^2),
\end{equation}
where $R_I$ indicates the depth of the bulk flow measurement. In this
work, we adopt $R_I = $ 20\hMpc, 30\hMpc~and 40\hMpc, in order to show
the bulk flow across a range of depths.

Since the 2MTF sample has a denser galaxy distribution in the sky
north of $\delta = -40^\circ$, we introduced the weight $w^d_i$ to
correct the uneven number density in these two sky areas. 
The regions of the sky north and south of this line contain 1,827 and 191 
2MTF galaxies and subtend 33,885 and 7,368 deg$^2$, respectively. This 
yields a ratio of galaxy number density between the southern and northern 
regions of roughly 1:2.08. We therefore set $w_i^d=2.08$ for the southern 
objects and $w_i^d=1$ for the northern ones.
%We set 
%$\delta = -40^\circ$ as a limit line, this line splits the sphere into 2 parts,
%the areas of these two parts are 7,368 deg$^2$ and 33,885 deg$^2$
%respectively. In the region south of $-40^\circ$, there are
%191 2MTF galaxies. 1,827 galaxies are located in the
%northern part. The ratio of the galaxy number density between southern
%and northern sky is roughly 1:2.08. To make the weighted number
%density in 2 sky regions are approximately the same, we therefore gave the
%southern galaxies a weight $w^d_i = 2.08$, and the weight of northern
%galaxies as $w^d_i = 1$.

To reduce the effect of outliers, we applied two cuts during the
fitting process.  With each successive cut, we did the $\chi^2$
minimization first, and compared the difference between the bulk
flow-corrected logarithmic distance ratio and the predicted Hubble
flow logarithmic distance ratio: $\Delta_i =\log(d_{z,i} / d_{\mathrm{model},
  i})- \log(d_{z,i} / d_{\mathrm{TF}, i})$.  The outliers with $\Delta_i >
3\sigma_i$ were then removed. We did the clipping twice, then applied
the third $\chi^2$ minimization, and report the result as the final
bulk flow velocity. The total number of galaxies removed from the sample 
is around 60.

The fitting errors were calculated via the jackknife method. We built
50 jackknife subsamples by randomly removing 2\% of the 2MTF sample,
ensuring that each galaxy was removed in one subsample only.  For every
jackknife subsample, the $\chi^2$ minimization fitting was adopted,
and the error on the bulk flow was taken as:
\begin{equation}
\epsilon_{V}=\left[ \frac{N-1}{N} \sum^{N}_{i=1} \left(V_{i}^{J}-\overline{V}^J \right)^2 \right]^{1/2},
\label{eq:error}
\end{equation}
where $N = 50$ is the number of jackknife subsamples, $V_{i}^{J}$ is
the bulk flow for the $i$th jackknife subsample, and $\overline{V}^J =
\frac{1}{N} \sum^{N}_{i=1} V_{i}^{J}$. The velocity components in three
directions ($V_X, V_Y, V_Z$) were estimated using Equation~\ref{eq:error}.

To analyze the performance of the $\chi^2$ minimization method, 
we simulated 1,000 mock catalogs and tested the method with these simulated 
data. We built the mock catalogs based on the real 2MTF data sample, with each 
catalog containing 2,018 mock galaxies.  These galaxies have the same sky position and redshift as the real 2MTF 
galaxies. 
%In each mock catalog, we set the bulk flow velocity to a random direction and 
%a random amplitude in the region of $(0^\circ \leq l \leq 360^\circ, -90^\circ \leq b \leq 90^\circ)$ 
%and 200\ \kms $\leq V \leq $ 400\ \kms, respectively. 
In each mock catalog, we set the input bulk flow velocity equals to the same 
value as our measurement of the K-band data sample at the depth 
of $R_I = $ 30\hMpc, i.e., $V_X = 163.4$ \kms, 
$V_Y = -308.3$ \kms\ and $V_Z = 107.7$ \kms.
We simulated distance errors by using a Gaussian random variable to 
induce mock scatter in the TF, with a variance equals to the total logarithmic 
distance error for that galaxy.  The average value of this logarithmic 
distance error is 0.096, as in the real data.

\begin{figure*}
\centering
\includegraphics[width=0.6\columnwidth, angle=-90]{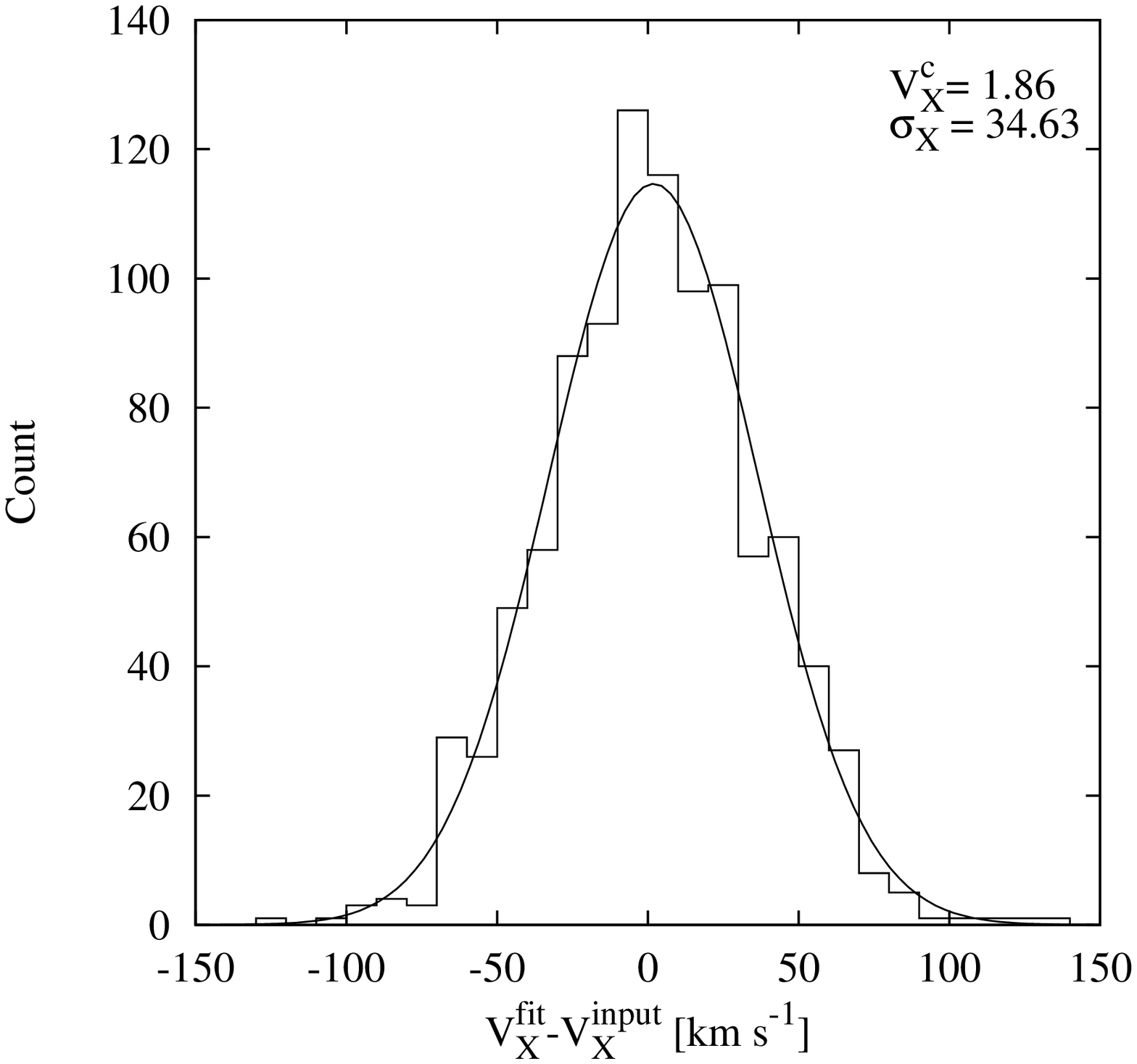}\hspace*{\fill}
\includegraphics[width=0.6\columnwidth, angle=-90]{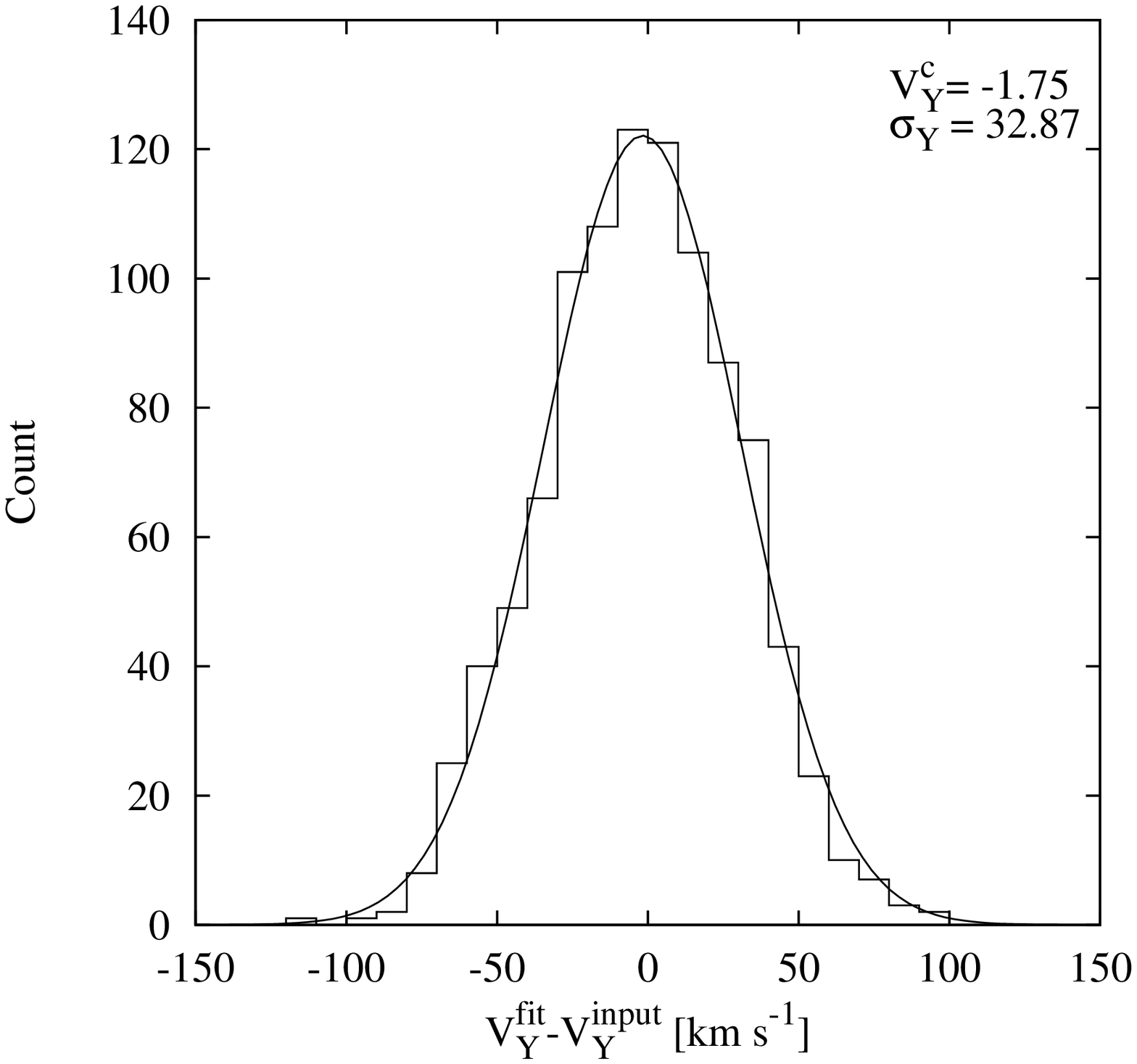}\hspace*{\fill}
\includegraphics[width=0.6\columnwidth, angle=-90]{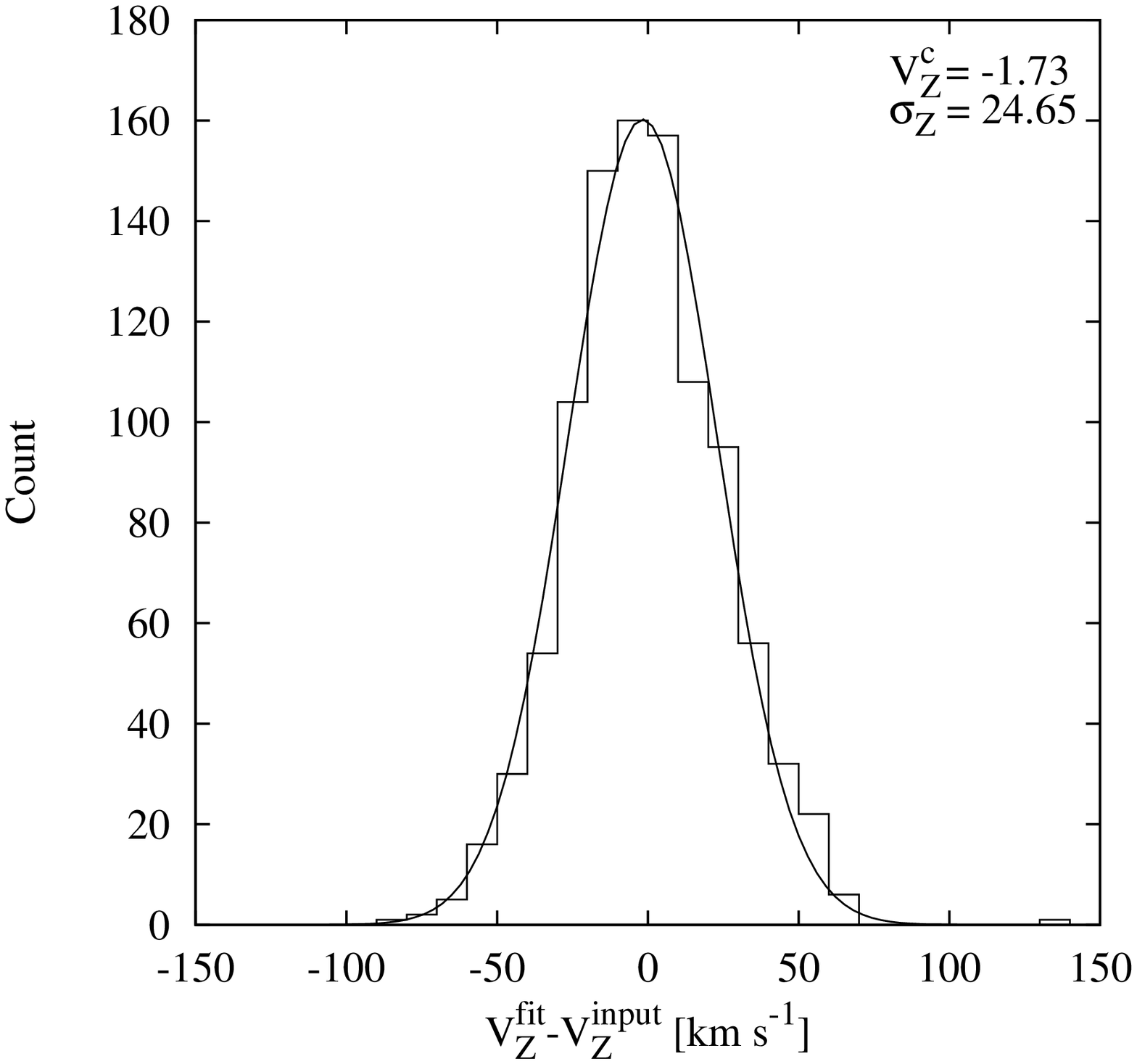}\hspace*{\fill}
\caption{Histograms of the differences between the input and output bulk flow 
components $V_X, V_Y$ and $V_Z$ for our 1,000 mock samples. 
The solid lines show the best fit Gaussians to the 
distributions. The Gaussian centers are located at $V_X^c = 1.86$ \kms, $V_Y^c=-1.75$ \kms and $V_Z^c = -1.73$ \kms\ with 
standard deviations of $\sigma_X = 34.63$ \kms, $\sigma_Y = 32.87$ \kms\ and 
$\sigma_Z = 24.65$ \kms\ respectively.}
\label{fig:mock}
\end{figure*}

We fit the bulk flow for each of the 1,000 mock catalogs using the $\chi^2$ minimization method, and plot the 
histograms of differences between the input and output bulk flow components $V_X, V_Y$ and $V_Z$ in Figure~\ref{fig:mock}. We then fit the histograms using 
Gaussian functions, and found the distribution of 1,000 `observed' bulk flow components 
centered around the input 
values with a scatter of $\sigma_{X} \sim 35$ \kms, $\sigma_{Y} \sim 33$ \kms\ and $\sigma_{Z} \sim 25$ \kms\  respectively. The distributions present very small shifts of $\sim$ 2 \kms\ in three components.  
This is smaller than our quoted bulk flow error by roughly an order of magnitude, and shows that the 
$\chi^2$ minimization method provides an unbiased fit of the bulk flow motion.

The best fit $\chi^2$ bulk flow results are presented in
Table~\ref{tab:bulk}, subdivided by wavelength and measuring depth $R_I$. 
In all three bands with three different depths,
we detected a non-zero bulk flow with a confidence level of at least
$3\sigma$.

In addition to the bulk flow measurements for individual wavebands, we
have also combined the data from all three wavebands into a composite
measurement of the bulk flow.  That is, each galaxy appears in this
sample three times, but with different peculiar velocities as measured
in each of the wavebands. But we have applied the $\chi^2$
minimization method so that each of the three peculiar velocity
measurements for a given galaxy is counted separately in Equation~\ref{eq:chi2}.
We then compute the error using the same jackknife method.  Though
once a galaxy was removed from a subsample, all three peculiar
velocities belonging to this galaxy were removed together.  We report
these composite measurements of the bulk flow as `3 bands combined' in
Table 1.  
%We note that because the magnitudes in the three bands are
%all heavily correlated, the errors on the composite measurement are
%very similar to the K band measurement, which has the smallest
%peculiar velocity errors of the three bands.

%
\begin{table*}
\caption[]{Best fit $\chi^2$ measured bulk flow of the 2MTF sample in
  J, H and K bands. The velocity components $(V_X, V_Y, V_Z)$ are
  calculated in Galactic Cartesian coordinates.}
\label{tab:bulk}
\centering
\begin{tabular}{lccccccc}
\hline
\hline
&Amplitude & $l$ & $b$ &$V_X$&$V_Y$&$V_Z$&$\chi^2/\rm{d.o.f.}$\\
 & \kms & deg & deg &\kms&\kms&\kms\\
\hline
\hline
J band \\
\hline
$R_I=20$~\hMpc & $303.5 \pm 22.2$& $306.7 \pm 7.3$& $11.9 \pm 4.9$& $177.4 \pm 29.2$& $-238.2 \pm 17.0 $& $62.5 \pm 22.2$ & 0.98\\ 
%$\uparrow $ the error of this lineis not reliable now.\\
$R_I=30$~\hMpc & $319.0 \pm 28.0$& $297.5 \pm 4.1$& $18.5 \pm 4.6$& $139.6 \pm 17.5$& $-268.4 \pm 27.9$& $101.1 \pm 26.9$ & 0.95\\ 
$R_I=40$~\hMpc & $319.4 \pm 21.3$& $290.8 \pm 4.2$& $11.4 \pm 4.7$& $111.3 \pm 20.9$& $-292.7 \pm 19.8$& $63.0 \pm 25.6$ & 0.95\\ 
\hline
\hline
H band \\
\hline
$R_I=20$~\hMpc & $345.7 \pm 53.6$& $308.7 \pm 7.8$& $18.1 \pm 13.9$& $205.7 \pm 54.4$& $-256.3 \pm 43.5 $& $107.1 \pm 82.2$ & 1.07\\ 
$R_I=30$~\hMpc & $352.4 \pm 34.4 $& $296.9 \pm 5.6$& $26.8 \pm 11.6$& $142.4 \pm 29.6$& $-280.6 \pm 31.7$& $158.7 \pm 65.2$ & 1.04\\ 
$R_I=40$~\hMpc & $319.1 \pm 37.4$& $296.2 \pm 8.3$& $5.1 \pm 6.1$& $140.5 \pm 51.1$& $-285.0 \pm 28.2$& $28.4 \pm 33.6$ & 1.00\\ 
\hline
\hline
K band \\
\hline
$R_I=20$~\hMpc & $301.3 \pm 27.3$& $305.4 \pm 8.2$& $8.5 \pm 6.4$& $172.6 \pm 42.1$& $-243.0 \pm 30.5$& $44.4 \pm 39.6$ & 1.00\\ 
$R_I=30$~\hMpc & $365.1 \pm 36.2$& $297.9 \pm 6.1$& $17.2 \pm 5.5$& $163.4 \pm 36.8$& $-308.3 \pm 44.5$& $107.7 \pm 33.8$ & 0.98\\ 
$R_I=40$~\hMpc & $331.1 \pm 22.5 $& $292.0 \pm 3.4$& $11.8 \pm 4.2$& $121.4 \pm 25.0$& $-300.5 \pm 19.6$& $67.9 \pm 20.8$ & 0.96\\ 
\hline
\hline
3 bands combined \\
\hline
$R_I=20$~\hMpc & $310.9 \pm 33.9$& $287.9 \pm 5.9$& $11.1 \pm 3.4$& $93.7 \pm 34.4$& $-290.4 \pm 28.9$& $59.8 \pm 18.4$ & 0.90\\ 
$R_I=30$~\hMpc & $280.8 \pm 25.0$& $296.4 \pm 16.1$& $19.3 \pm 6.3$& $117.8 \pm 68.3$& $-237.4 \pm 30.6$& $92.9 \pm 29.4$ & 0.93\\ 
$R_I=40$~\hMpc & $292.3 \pm 27.8 $& $296.5 \pm 9.8$& $6.5 \pm 9.2$& $129.6 \pm 47.6$& $-259.9 \pm 37.6$& $33.0 \pm 46.6$ & 0.95\\ 
\hline
\end{tabular}
\end{table*}

\subsection{Additional bulk flow fitting methods}

In addition to the $\chi^2$ minimization method used in Section~\ref{sec:method},
we have done additional fits of the bulk flow using the maximum
likelihood estimator (MLE) \citep{Kaiser1988} and the minimum variance
method (MV) \citep{Watkins2009, Feldman2010}.  Both methods involve
converting the logarithmic distance ratios into linear peculiar
velocities, and then applying the appropriate weights, which depend on
sky position. We thus have
\begin{equation}
\overrightarrow{V} = \sum_{n=1}^{N} w_{n} v_n,
\end{equation}
where $v_n$ is the linear peculiar velocity of galaxy $n$, and $w_{n}$
is the corresponding weight applied to the galaxy such that one
recovers the bulk flow $\overrightarrow{V}$.  The optimal approach for
converting the logarithmic distance ratios into linear peculiar
velocities for use in these methods is explained in detail by
Scrimgeour et al. (in prep.), who apply these methods to the 6dF
Galaxy Survey peculiar velocities (Springob et al., submitted). 
The advantage of these methods
over the $\chi^2$ approach described in Section~\ref{sec:method} is that the
weighting accounts for the sample selection on the sky, with more
densely sampled regions being weighted differently from less densely
sampled regions.  The advantage of the $\chi^2$ minimization approach,
however, is that the fitting is done in the logarithmic distance units
in which the errors are roughly Gaussian.

\subsubsection{Maximum Likelihood Method}

Following \citet{Kaiser1988}, the MLE bulk flow weights are
\begin{equation}
w_{i,n}=\sum_{j} A_{i,j}^{-1} \frac{\widehat{r}_{n,j}}{\sigma_{n}^{2} + \sigma_{*}^{2}},
\end{equation}
where
\begin{equation}
A_{i,j}=\sum_{n} \frac{\widehat{r}_{n,i} \widehat{r}_{n,j}}{\sigma_{n}^{2} + \sigma_{*}^{2}},
\end{equation}
where $\sigma_n$ is the uncertainty on the velocity $v_n$ which is
calculated following Scrimgeour et al. (in prep.), and $\sigma_*$ is
the 1D velocity dispersion.

\subsubsection{Minimum Variance Method}

The Minimum Variance weighting scheme, as proposed by
\citet{Watkins2009} and \citet{Feldman2010}, gives us an
$N$-dimensional vector of weights specifying the $i$th moment $u_i$,
which are given by:
\begin{equation}
\overrightarrow{w_i} = ({\mathbf G}+\lambda {\mathbf P})^{-1} {\mathbf Q_i}
\end{equation}
where $\overrightarrow{w_i}$ is the vector of weights for each galaxy
in the $i$th direction.  {\bf G} is the covariance matrix of the
individual velocities, which includes both a noise term and a cosmic
variance term determined by an input model power spectrum.  {\bf P} is
the $k=0$ limit of the angle-averaged window function of the galaxies,
and ${\mathbf Q_i}$ incorporates the input ideal window function. The
$\lambda$ term is a Lagrange multiplier. The MV method weights the
bulk flow measurements to a given ideal survey, which we have chosen
to be the same Gaussian profile with the $\chi^2$ minimization method
(Equation~\ref{eq:density} and \ref{eq:number_dist}) with $R_I =$ 20,
30 and 40\hMpc. See \citet{Watkins2009} and \citet{Feldman2010} for
more details on the method.

The resulting MLE and MV measurements of the bulk flow are presented
in Table 2, and they largely agree with the $\chi^2$ minimization
method.  We discuss these results further in the following section.

\begin{table*}
\caption[]{Minimum Variance (MV) and Maximum Likelihood Estimate (MLE)
  bulk flow of the 2MTF sample in J, H and K bands. The errors quoted
  are noise-only, with noise+cosmic variance in parentheses.}
\label{tab:MV}
\centering
\begin{tabular}{lcccccc}
\hline
\hline
&Amplitude & $l$ & $b$ &$V_X$&$V_Y$&$V_Z$\\
 & \kms & deg & deg &\kms&\kms&\kms\\
\hline
\hline
J band \\
\hline
MV ($R_I=20$~\hMpc) & $313 \pm 33(150)$& $295 \pm 7$& $13 \pm 6$& $129 \pm 35(174)$& $-276 \pm 33(173) $& $71 \pm 30(172)$\\ 
%$\uparrow $ the error of this lineis not reliable now.\\
MV ($R_I=30$~\hMpc) & $328 \pm 39(137)$& $283 \pm 7$& $12 \pm 6$& $72 \pm 39(151)$& $-313 \pm 39(153)$& $68\pm 36(150)$\\ 
MV ($R_I=40$~\hMpc) & $325 \pm 49(132)$& $275 \pm 9$& $7 \pm 8$& $27 \pm 48(144)$& $-321 \pm 50(147)$& $41 \pm 45(142)$\\ 
MLE & $351 \pm 28$& $295 \pm 5$& $17 \pm 4$& $143 \pm 30$& $-304 \pm 28$& $100 \pm 21$\\ 
\hline
\hline
H band \\
\hline
MV ($R_I=20$~\hMpc) & $294 \pm 33(149)$& $294 \pm 7$& $13 \pm 6$& $177 \pm 35(173)$& $-261 \pm 33(172) $& $67 \pm 30(173)$\\ 
MV ($R_I=30$~\hMpc) & $317 \pm 39(135) $& $283 \pm 7$& $14 \pm 7$& $70 \pm 39(151)$& $-300 \pm 39(153)$& $75 \pm 36(150)$\\ 
MV ($R_I=40$~\hMpc) & $308 \pm 48(132)$& $275 \pm 9$& $10 \pm 8$& $29 \pm 48(144)$& $-302 \pm 49(147)$& $55 \pm 45(142)$\\ 
MLE & $338 \pm 28$& $294 \pm 5$& $18 \pm 4$& $133 \pm 30$& $-292 \pm 28$& $104 \pm 21$\\ 
\hline
\hline
K band \\
\hline
MV ($R_I=20$~\hMpc) & $305 \pm 33(149)$& $292 \pm 7$& $13 \pm 6$& $109 \pm 34(172)$& $-276 \pm 33(174)$& $71 \pm 29(173)$\\ 
MV ($R_I=30$~\hMpc) & $313 \pm 39(134)$& $276 \pm 7$& $13 \pm 7$& $33 \pm 39(151)$& $-303 \pm 40(153)$& $70 \pm 36(149)$\\ 
MV ($R_I=40$~\hMpc) & $312 \pm 49(132) $& $270 \pm 9$& $11 \pm 8$& $1 \pm 48(145)$& $-306 \pm 50(147)$& $59 \pm 45(142)$\\ 
MLE & $345 \pm 27$& $295 \pm 5$& $16 \pm 4$& $141 \pm 30$& $-300 \pm 28$& $95 \pm 21$\\ 
\hline

\end{tabular}
\end{table*}

\section{Discussion}
\label{sec:compare}
In Table~\ref{tab:bulk} we list 12 $\chi^2$ minimization measured bulk
flow velocities from the three 2MASS near infrared wavebands and the
`3 bands combined' sample, each using three sample depths.  To
simplify the discussion here, we focus on our best measurements of the
bulk flow at the three depths, which are the measurements taken from
the combined sample.
%We choose the K-band results mainly because 
%two reasons, firstly, the K-band Tully-Fisher relation provides smaller intrinsic 
%errors than H and J bands, the second reason is that the bulk flow fitting results
%in K band sample have better reduced Chi-square values.

%\subsection{The direction of bulk flow velocities}
Using the combined data with a depth $R=30$\hMpc, we detected a bulk
flow velocity in the direction of $l =296.4^\circ \pm 16.1^\circ, b =
19.3^\circ \pm 6.3^\circ$.  Our direction is consistent with that of
most previous studies, but our result tends to have a higher galactic
latitude.

\begin{figure*}
\centering
\includegraphics[width=1.1\columnwidth, angle=-90]{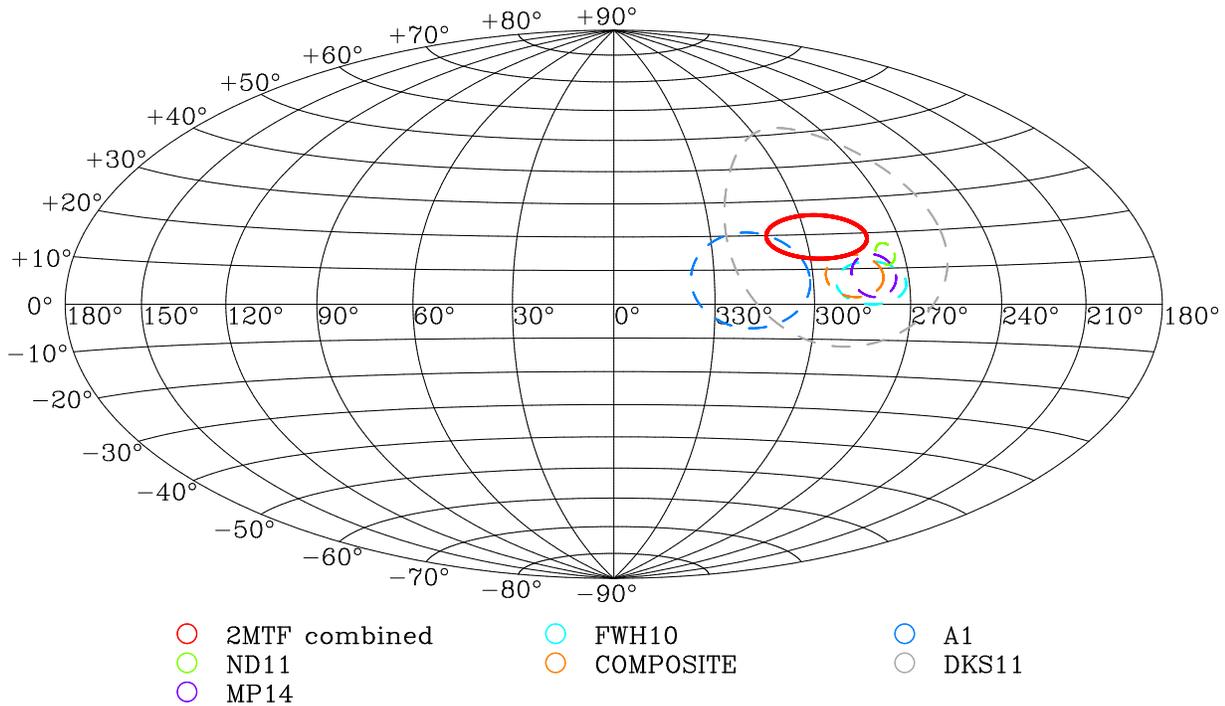}
\caption{The direction of the measured bulk flow velocities, using an 
Aitoff projection and Galactic coordinates. The red circle shows the bulk flow
  direction estimated from `3 bands combined' sample with depth $R_I =
  30$\hMpc. The size of the circles indicates the
  $1\sigma$ error of the direction. The bulk flow directions from
  several previous studies are also plotted for comparison. The
  references for these literature results are listed in
  Table~\ref{tab:direction}.}
\label{fig:bf_dir}
\end{figure*}

We list our $\chi^2$ minimization measured bulk flow directions, along
with the results from previous studies in Table~\ref{tab:direction}.
An Aitoff projection of bulk flow directions is also presented in
Figure~\ref{fig:bf_dir}.
\begin{table}
\caption[]{Bulk flow directions from previous studies}
\label{tab:direction}
\centering
\begin{tabular}{lccc}
\hline
\hline
 & $l$ & $b$ & Reference\\
 & deg & deg &\\
\hline
\bf 2MTF &  $\bf 296.4 \pm 16.1$ & $\bf 19.3 \pm 6.3$ & \bf This work\\
COMPOSITE & $ 287 \pm 9$ & $ 8 \pm 6$ & \citet{Watkins2009}\\
FWH10 & $ 282 \pm 11$ & $ 6 \pm 6$ & \citet{Feldman2010}\\
ND11 & $ 276 \pm 3$ & $ 14 \pm 3$ & \citet{Nusser2011}\\
DKS11 & $ 290 ^{+39}_{-31}$ & $ 20 ^{+32}_{-32}$ & \citet{Dai2011}\\
A1 & $ 319 \pm 25$ & $ 7 \pm 13$ & \citet{Turnbull2012}\\
MP14 & $ 281 \pm 7$ & $ 8 ^{+6}_{-5}$ & \citet{Ma2014}\\
\hline
\end{tabular}
\end{table}

%\subsection{The amplitude of bulk flow velocities}
The expected amplitude of the bulk flow is strongly dependent on the
characteristic depth of the galaxies being sampled
\citep{Li2012,Ma2014}.  As survey geometry and source selection
criteria vary greatly between surveys, one must account for this in
comparing the bulk flow amplitude to cosmological expectations.

In the $\Lambda$CDM model, the variance 
of the bulk flow velocity in a spherical region R is
\begin{equation}
v_{\mathrm{rms}}^2=\dfrac{H_0^2 f^2}{2\pi^2}\int W^2(kR)P(k)\textrm{d}k,
\end{equation}
where $k$ is the wavenumber, $W(kR) = \exp(-k^2R^2)/2$ is the Fourier
transform of a Gaussian window function, $P(k)$ is the matter power
spectrum \citep[we used the matter power spectrum generated by the
  CAMB package,][]{Lewis2000}, $f=\Omega_\mathrm{m}^{0.55}$ is the linear
growth rate, and $H_0$ is the Hubble constant.

The probability distribution function of the bulk flow V is
\citep{Li2012}
\begin{equation}
p(V)\textrm{d}V=\sqrt{\dfrac{2}{\pi}}\left(\dfrac{3}{v_{\mathrm{rms}}^2}\right)^{3/2}V^2
\exp\left(-\dfrac{3V^2}{2v_{\mathrm{rms}}^2}\right)\textrm{d}V,
\end{equation}
The PDF here is normalized. By setting $\textrm{d}p(V)/\textrm{d}V=0$,
the peak of the distribution can be easily calculated: $v_{\mathrm{peak}} =
\sqrt{2/3}v_{\mathrm{rms}}$. We adopt this peak velocity as the bulk velocity
amplitude predicted by the $\Lambda$CDM model.

\begin{figure}
\centering
\includegraphics[width=0.7\columnwidth, angle=-90]{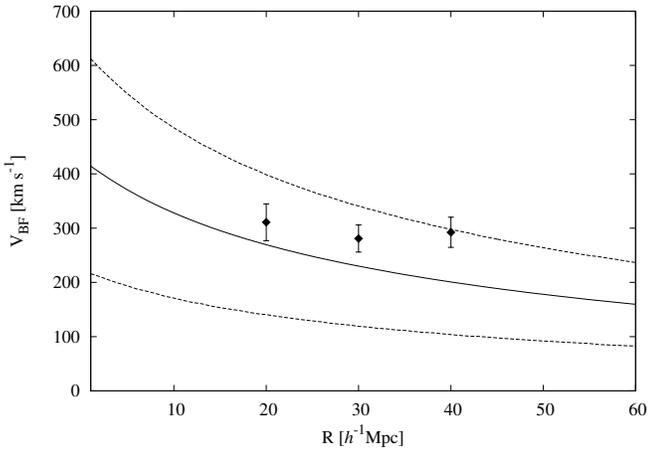}
\caption{The comparison between the bulk flow velocity amplitude from
  the 2MTF sample and the $\Lambda$CDM prediction using the WMAP-7yr 
  parameters \citep{Larson2011}. The
  diamonds with error bars indicate the bulk flow velocity amplitude
  measured from the 2MTF `3 bands combined' sample using the $\chi^2$
  minimization method, with the depth $R_I$ = 20, 30 and 40
  \hMpc\ respectively. The solid line shows the theoretical curve and
  the dashed lines indicate the sample variance at the $1\sigma$ level.}
\label{fig:theo}
\end{figure}

We show the 3 bands combined measurement of $\chi^2$ minimization
method in Figure~\ref{fig:theo} together with the model predicted
curve.  The $1\sigma$ variance of the bulk flow velocity is also
plotted as a dashed line. Our results agree with the model predicted
bulk flow amplitude at the $1\sigma$ level. Thus, our results support
the conclusions reported by some previous studies
\citep[e.g.][]{Dai2011,Nusser2011,Turnbull2012,Rathaus2013,Ma2014}
that the bulk flow detected in the local Universe is consistent with
the $\Lambda$CDM model.

\begin{figure*}
\centering
\includegraphics[width=1.1\columnwidth, angle=-90]{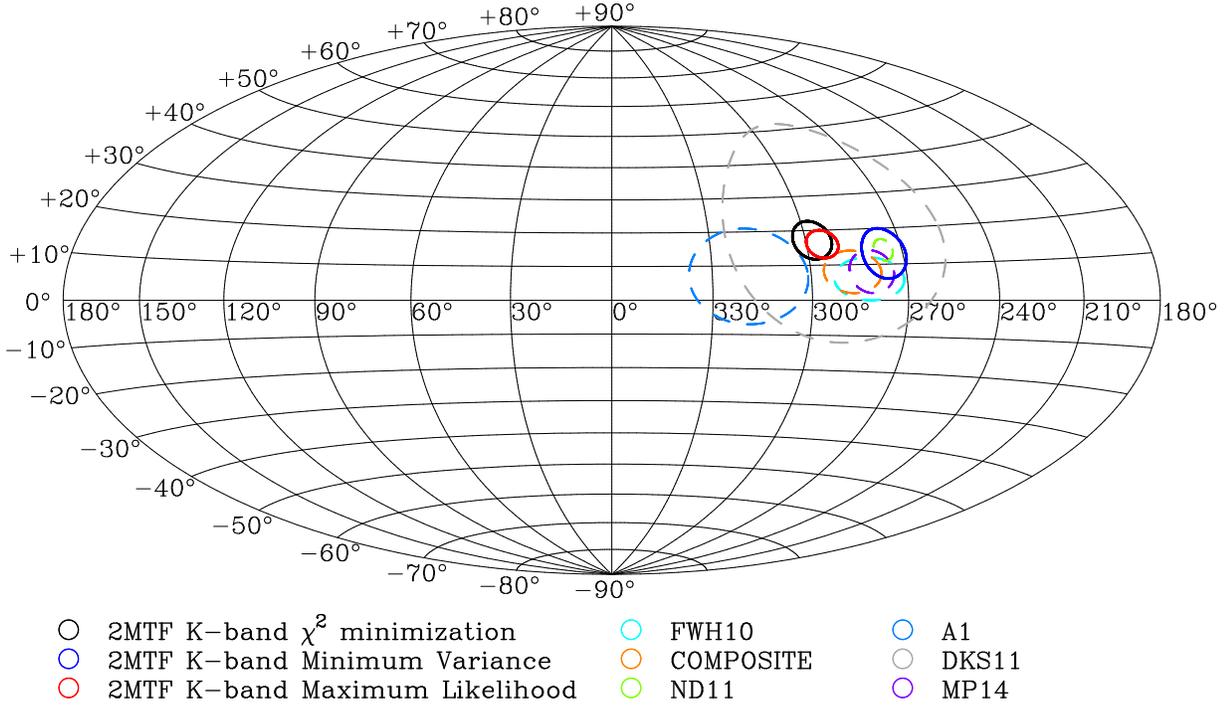}
\caption{The direction of the measured bulk flow velocities, using an 
Aitoff projection and Galactic coordinates. The black, blue, and red circles show the bulk
  flow direction determined from the K-band sample by the $\chi^2$
  minimization ($R_I=30$\hMpc), Minimum Variance ($R_I=30$\hMpc), and
  Maximum Likelihood methods respectively.  The size of the circles
  indicates the $1\sigma$ error.  The literature bulk flow directions
  are plotted by dashed circles.}
\label{fig:compare_direction}
\end{figure*}

\begin{figure}
\centering
\includegraphics[width=0.7\columnwidth, angle=-90]{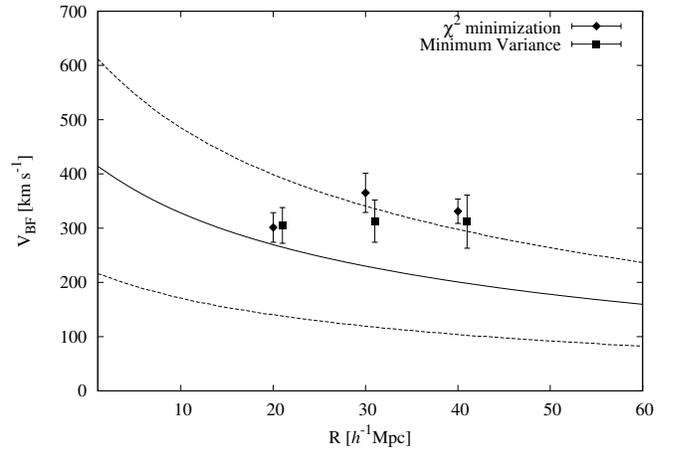}
\caption{The comparison between the bulk flow amplitudes measured by
  $\chi^2$ minimization method (diamonds) and minimum variance method
  (squares) on 2MTF K-band sample. The minimum variance method results
  are shifted 1\hMpc\ right to make the comparison clear. The solid
  line shows the theoretical curve and the dashed lines indicate the
  variance at the $1\sigma$ level.}
\label{fig:compare_amplitude}
\end{figure}

As a point of comparison, both the MV and MLE results are shown in
Table 2.  The MV calculation is performed for the same 20, 30, and 40
\hMpc\ depths used for the $\chi^2$ minimization method, while the MLE method
only samples one depth.  The measurements largely agree with the $\chi^2$ minimization 
results in both amplitude and direction.  A comparison of
bulk flow measurements on 2MTF K-band sample is shown in Figures
\ref{fig:compare_direction} and \ref{fig:compare_amplitude}.

\section{Conclusions}
\label{sec:conclu}
The bulk flow motion is the dipole component of the peculiar velocity
field, thought to be induced by the gravitational attraction of
large-scale structures. Therefore, its measurement can help to constrain
cosmological models.

The 2MTF survey is an all-sky Tully-Fisher survey, using photometric
data from the 2MASS catalog and \HI\ data from existing published works
and new observations.  Using this dataset, we have derived peculiar
velocities for 2,018 galaxies in the local Universe ($cz \leq 10,000$
\kms).  This sample covers all of the sky down to a Galactic latitude
$|b| = 5^\circ$, providing better sky-coverage than previous surveys.

We applied a $\chi^2$ minimization method to the peculiar velocity
catalog to estimate the bulk flow motion. We applied three Gaussian
window functions located at three different depths ($R_I = $ 20\hMpc,
30\hMpc, and 40\hMpc) to the catalogs generated by three near-infrared
bands (J, H \& K). Our tightest constraints came
from combining the data in all three bands, which gave us bulk flow
amplitudes of $V = 310.9 \pm 33.9$ \kms, $V = 280.8 \pm 25.0$ \kms,
and $V = 292.3 \pm 27.8$ \kms\ for $R_I = $ 20\hMpc, 30\hMpc\ and
40\hMpc\ respectively.  Similar results are found when we apply the
maximum likelihood and minimum variance methods of \citet{Kaiser1988}
and \citet{Watkins2009} respectively.  We find that these amplitudes
all agree with the $\Lambda$CDM model prediction at the $1\sigma$
level.  The directions of our estimated bulk flow are also consistent
with previous probes.\\

The authors gratefully acknowledge Martha Haynes, Riccardo Giovanelli,
and the ALFALFA team for supplying the latest ALFALFA survey data. We
thank Yin-zhe Ma for useful comments and discussions.

The authors wish to acknowledge the contributions of John Huchra (1948 - 2010) 
to this work. The 2MTF survey was initiated while KLM was a post-doc working 
with John at Harvard, and its design owes much to his advice and insight. This 
work was partially supported by NSF grant AST- 0406906 to PI John Huchra.

Parts of this research were conducted by the Australian Research
Council Centre of Excellence for All-sky Astrophysics (CAASTRO),
through project number CE110001020. TH was supported by the National
Natural Science Foundation (NNSF) of China (11103032 and 11303035) and
the Young Researcher Grant of National Astronomical Observatories,
Chinese Academy of Sciences.

\bibliographystyle{apj}
\bibliography{bibfile}

\begin{appendix}
\section{New intrinsic error estimation}
\label{sec:intrinsic}
As described in section~\ref{sec:photo}, \citet{Masters2008} estimated
the intrinsic errors of the Tully-Fisher relation using the 888-galaxy
template sample with I-band axis-ratios.  However, we have adopted the
2MASS co-added axis-ratio for the 2MTF sample.  Thus, we must make a
new estimate of the intrinsic scatter in the TF relation, appropriate
for the co-added axis ratios.  We have calculated the new intrinsic
errors by subtracting the observed error components from the total
scatter of the distribution.

We assume that a proper error estimation should approximately match 
the total scatter of the data sample: 
\begin{equation}
\label{eq:error2}
\sigma^2 \sim \epsilon_{\mathrm{total}}^2 = \epsilon_{\mathrm{wid}}^2+\epsilon_{\mathrm{ran}}^2
+\epsilon_{\mathrm{mag}}^2+\epsilon_{\mathrm{inc}}^2+\epsilon_{\mathrm{int}}^2,
\end{equation}
where $\sigma$ is the scatter of the data points related to the
Tully-Fisher template, $\epsilon_{\mathrm{wid}}$ is the error of the \HI\ 
widths, $\epsilon_{\mathrm{ran}} = 268$ \kms\ is the assumed mean rms of
velocities for field galaxies (see details in
Appendix~\ref{sec:random}), $\epsilon_{\mathrm{mag}}$ is the error on the 2MASS
magnitudes, $\epsilon_{\mathrm{inc}}$ is the error introduced by the 2MASS
co-added inclinations, $\epsilon_{\mathrm{int}}$ is the intrinsic error
of the Tully-Fisher relation.  All errors and scatter in
Equation~\ref{eq:error2} are in logarithmic units.

The residual component or intrinsic error is
\begin{equation}
\label{eq:leftover}
\epsilon_{\mathrm{int}}^2 \sim \sigma^2 - \left(\epsilon_{\mathrm{wid}}^2+\epsilon_{\mathrm{ran}}^2
+\epsilon_{\mathrm{mag}}^2+\epsilon_{\mathrm{inc}}^2\right).
\end{equation}

We assume a linear relation between the intrinsic error and
logarithmic \HI\ widths $\log W$
\begin{equation}
\label{eq:fitting}
\epsilon_{\mathrm{int}} = a\log W + b,
\end{equation}
where $a$ and $b$ are free parameters.

\begin{figure}
\centering
\includegraphics[width=0.7\columnwidth, angle=-90]{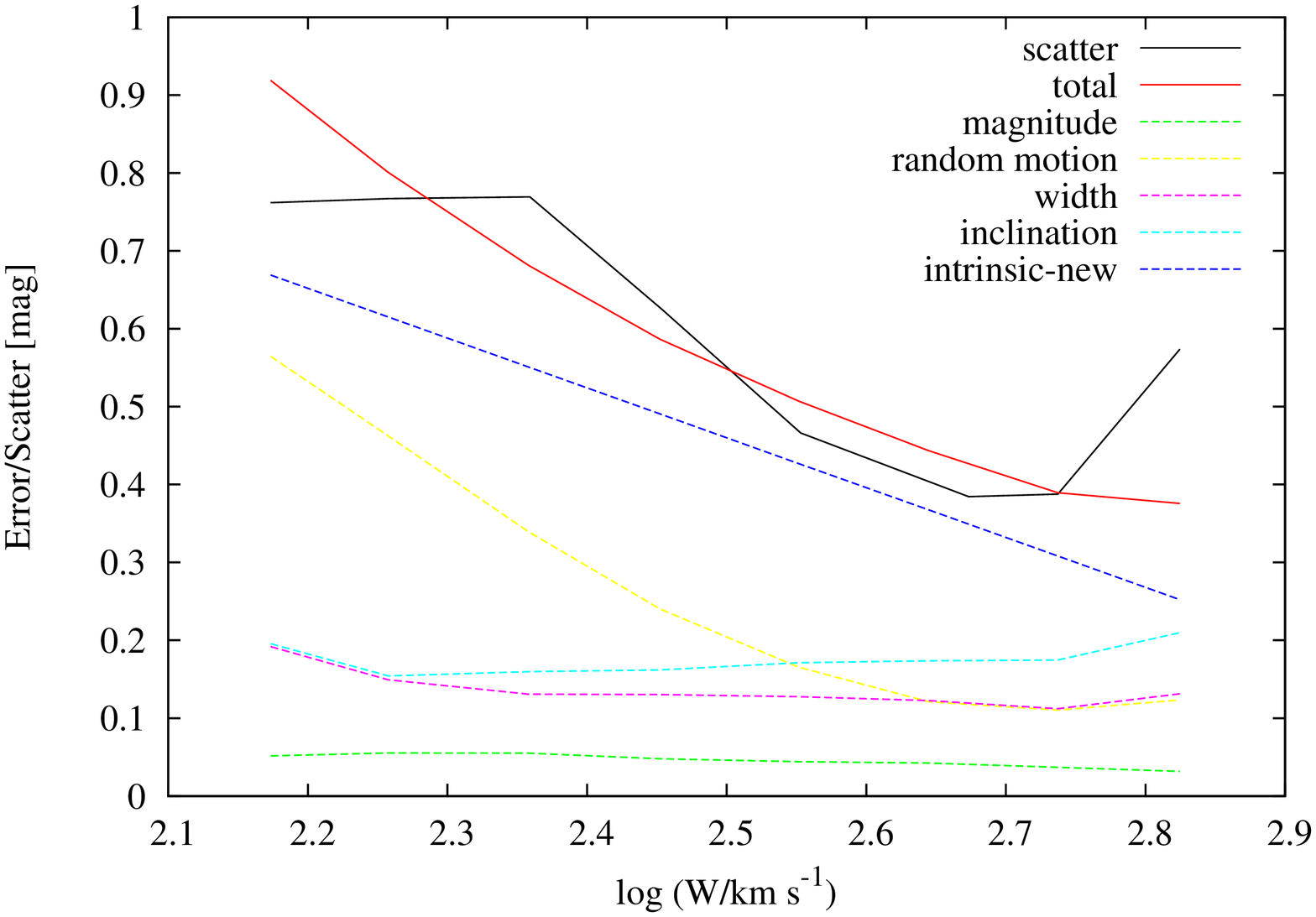}
\caption{A plot of different error components of the K-band data as a
  function of the logarithmic \HI\ width. The blue dashed line indicates
  the new intrinsic error term.  The
  red line shows our estimate of the final total error in the data, which matches the
  observed scatter of the sample well.}
\label{fig:intrinsic}
\end{figure}

The final results are the intrinsic error terms reported by
Equation~\ref{eq:intrinsic}. We show the error components along with
the total scatter of the K-band sample in the
Figure~\ref{fig:intrinsic}.  Using the new intrinsic error, the total
error closely matches the total scatter.
\section{The random motion of galaxies}
\label{sec:random}
Ideally, the intrinsic error of the Tully-Fisher relation would be
estimated using a galaxy cluster sample.  All of the galaxies in the
same cluster are assumed to be at the same distance, thus subtracting
out the individual motions of the galaxies themselves. However, when
estimating the new intrinsic error term, we used the data of 
2,018 field galaxies.  The uncorrected random motions of these
galaxies would introduce a spurious component into the scatter of the
Tully-Fisher relation.

\begin{figure}
\centering
\includegraphics[width=0.7\columnwidth, angle=-90]{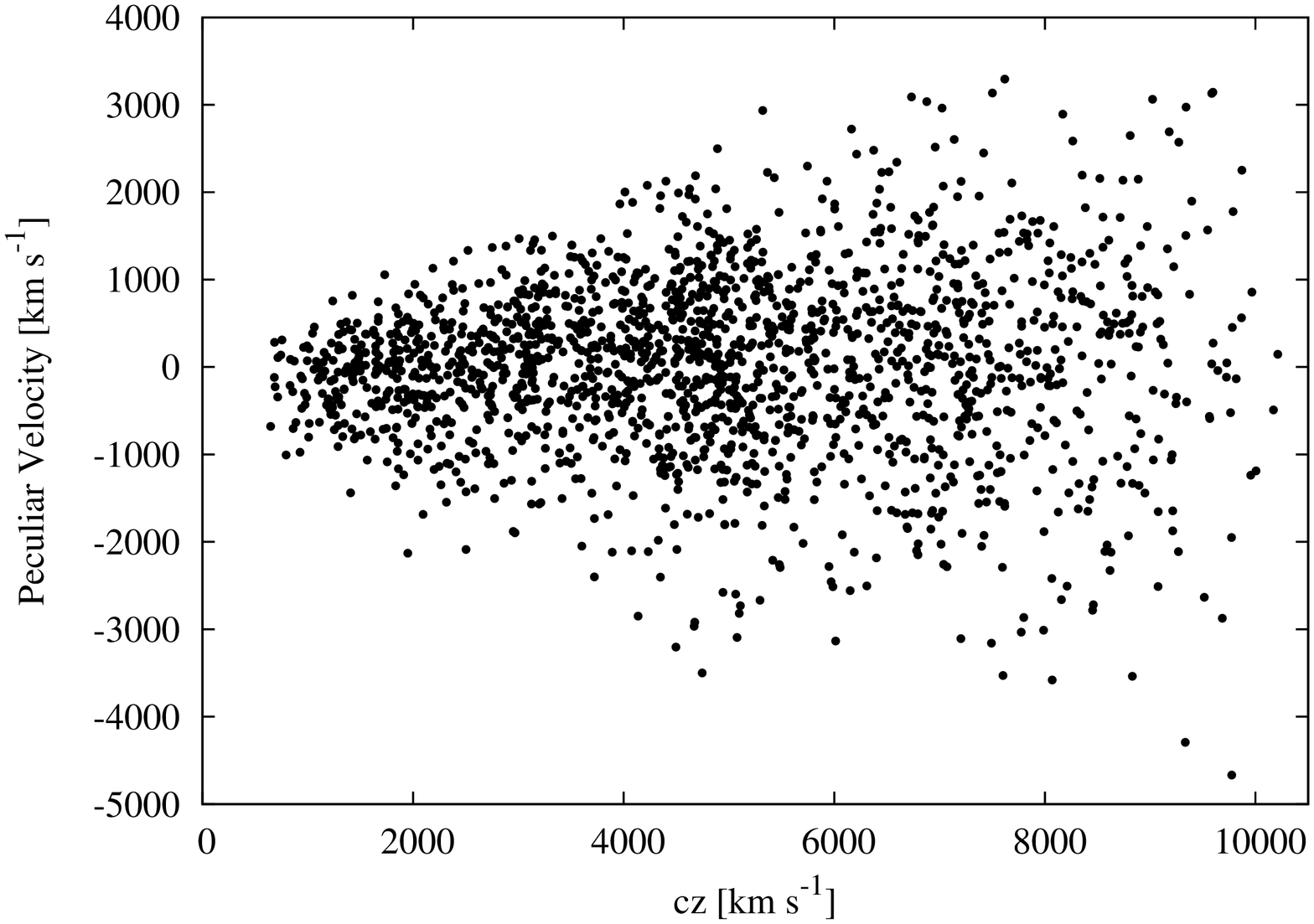}
\caption{Peculiar velocities vs. redshift for the 2,018 2MTF field
  galaxies.}
\label{fig:cz_pv}
\end{figure}

\begin{figure}
\centering
\includegraphics[width=0.7\columnwidth, angle=-90]{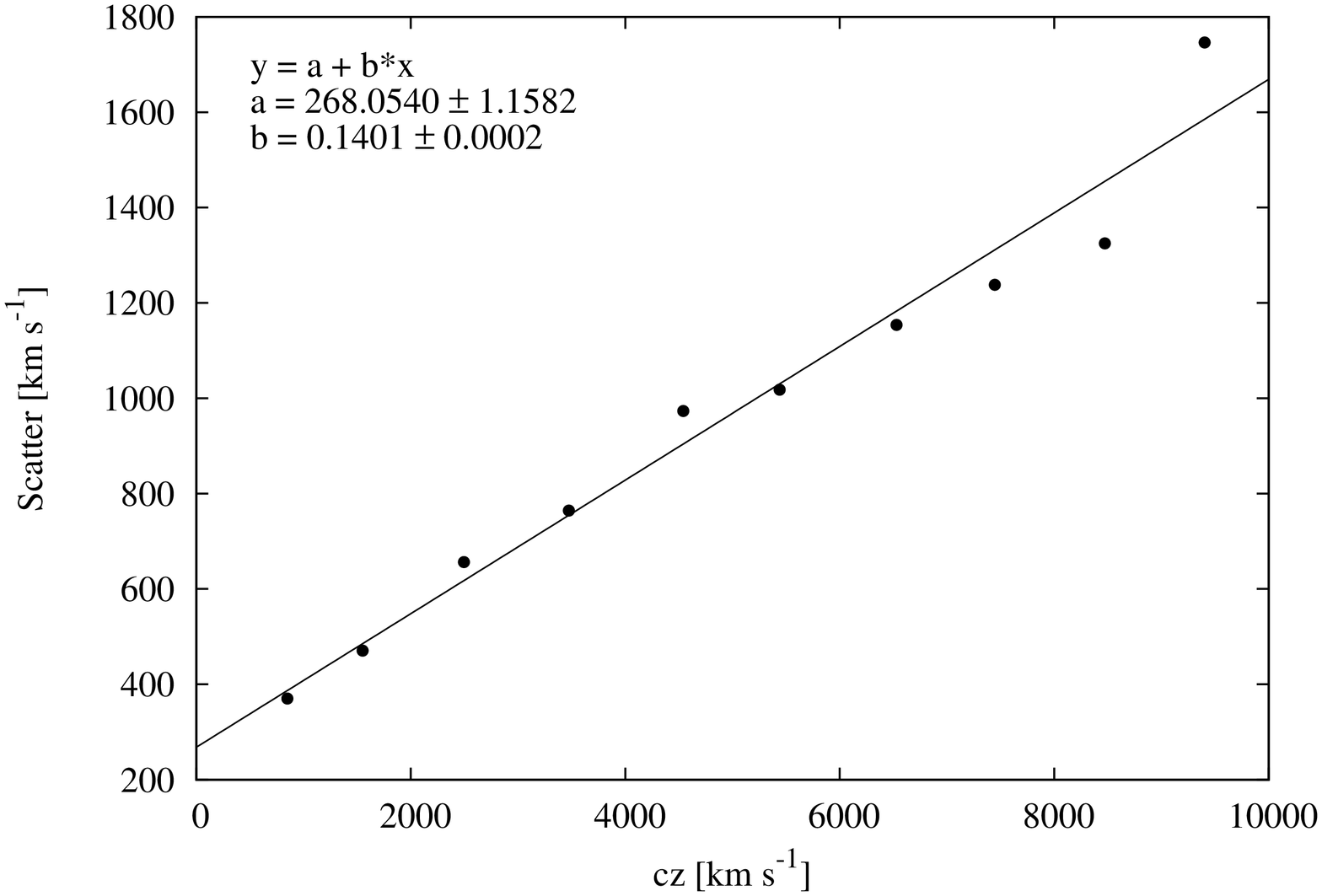}
\caption{Mean scatter of peculiar velocity as a function of redshift,
  with the best fit linear relation superimposed.}
\label{fig:ran_fit}
\end{figure}

To remove this effect, we need to estimate the mean random motion of
galaxies in our sample. We start by calculating the peculiar velocity
of the field galaxies using the Tully-Fisher relation template from
Equation~\ref{eq:tf}. Instead of the logarithmic quantity $\log(d_z/d_{\mathrm{TF}})$, 
a linear low redshift approximation 
\begin{equation}
v_{\mathrm{pec}}=cz\left(1-10^{\frac{\Delta M}{5}}\right),
\end{equation}
is adopted to generate the peculiar velocities of galaxies. 
We then place the galaxies onto the redshift -
peculiar velocity diagram Figure~\ref{fig:cz_pv}, and calculate 
peculiar velocity scatter in redshift bins of width 1000 \kms.  As
expected, the scatter increases with redshift, owing to the fact that
most of the observational errors scale with distance. To find the pure
random motion velocity of the galaxies, we do a linear fit to the
scatter as a function of redshift, and extended the linear relation to
$cz = 0$ \kms.  We adopt this zero point as the underlying rms of our
velocities.  As shown in Figure~\ref{fig:ran_fit}, we found $v_{\mathrm{ran}} =
268$ \kms, which is close to the commonly used value of $v_{\mathrm{ran}} =
300$ \kms \citep[e.g.,][]{Strauss1995, Masters2006}.

\end{appendix}

\label{lastpage}
\end{document}